# Understanding brain networks and brain organization

Luiz Pessoa
Department of Psychology and Maryland Neuroimaging Center
University of Maryland, College Park
March 23, 2014

**Abstract**

What is the relationship between brain and behavior? The answer to this question necessitates characterizing the mapping between structure and function. The aim of this paper is to discuss broad issues surrounding the link between structure and function in the brain that will motivate a network perspective to understanding this question. As others in the past, I argue that a network perspective should supplant the common strategy of understanding the brain in terms of individual regions. Whereas this perspective is needed for a fuller characterization of the mind-brain, it should not be viewed as panacea. For one, the challenges posed by the many-to-many mapping between regions and functions is not dissolved by the network perspective. Although the problem is ameliorated, one should *not* anticipate a *one*-to-*one* mapping when the network approach is adopted. Furthermore, decomposition of the brain network in terms of meaningful clusters of regions, such as the ones generated by community-finding algorithms, does not by itself reveal "true" subnetworks. Given the hierarchical and multi-relational relationship between regions, multiple decompositions will offer different "slices" of a broader landscape of networks within the brain. Finally, I described how the function of brain regions can be characterized in a multidimensional manner via the idea of diversity profiles. The concept can also be used to describe the way different brain regions participate in networks.

## 1. From areas to networks

Much has been written about the issue of localizability of mental processes, a problem that is at the core of neuroscience as a scientific discipline. Even a cursory look at the field reveals a continual swing of the pendulum between holistic and modular explanations (for an excellent account, see Shallice's book [1]).

The simplest way to conceptualize the relationship between a brain area and behavior is to assume a one-to-one mapping between an area and its function. For example, the primary visual cortex is linked to visual perception, or a set of more basic visual functions, such as "edge detection". Such an exercise becomes considerably less straightforward for more central areas (that is, farther from the sensory periphery), but we can imagine extending it throughout the brain. The end product of such a strategy would be a list of area–function pairs: $L = \{(A_1,F_1), (A_2,F_2),\ldots, (A_n,F_n)\}$. Brain areas might then be labeled as "perceptual", "motor", "cognitive", "emotional", "motivational", and so on, based on their purported functions and how they are envisioned to shape behavior. For instance, we could describe the amygdala as emotional given its contributions to fear conditioning, and the dorsal-medial PFC as cognitive given its role in the processing of response conflict [2].

Disregarding for now the thorny issue of what precisely is meant by "area" and "function", it is readily apparent that brain regions participate in many functions, and that many functions are carried out by many regions (Figure 1). For instance, the dorsal-medial PFC is important for a diverse range of cognitive operations, as well as for emotional processing. This region thus provides an example of an area involved in many functions, namely an instance of a one-to-*many* mapping. Conversely, both frontal and parietal regions participate in attentional and executive processes, illustrating the situation of multiple regions carrying out a related function, an instance of a *many*-to-one mapping.

--- Figure 1 ---

More generally, the mapping between structure and function is both *pluripotent* (one-to-many) and *degenerate* (many-to-one). Pluripotentiality means that the same structural configuration can perform multiple functions. Degeneracy refers to the ability of structurally different elements to perform the same function or yield the same output [3] – or to be able to complete a task. Notably, degeneracy should be distinguished from *redundancy,* which occurs when structurally identical elements perform the same function (as in "back-up" engineering systems). To the extent that pluripotentiality and degeneracy are accepted concerning the mind-brain[1], the combination of the two indicates that there are no "necessary and sufficient" brain systems. In particular, the existence of two or more degenerate systems that do *not* overlap precludes the existence of a single necessary area for a given function [7].

In the above discussion, I bypassed the difficult question of what constitutes a brain region and, even more challengingly, what constitutes a function. Clearly, structure-function relationships can be defined at multiple levels, from the precise (for instance, primary visual cortex is concerned with edge detection) to the abstract (for instance, primary visual cortex is concerned with visual perception), and structure-function relationships will depend on the specific level that is targeted. Some authors have suggested that, at some levels of description, a brain region does *not* have more than one function. For instance, the left posterior fusiform gyrus in temporal cortex, which has been implicated in the processing of word forms, animal structures, and so on, can be described by a single, more abstract label of "sensori-motor integration" [8]. Price and Friston suggest that whether a region can have more than one function depends on the level of the relationship, such that at a sufficiently abstract level, a region will have a single function – though note that for this notion to be useful, the abstractness has to be relatively limited, and not simply a vague description such as "cognitive function". Although the search for better conceptualizations of a region's functions is valuable, I propose below that the region level is *inadequate* to describe how brain structure is linked to mental function. More forcefully, understanding the structure-function mapping at the level of brain regions is unproductive because regions are not a meaningful *unit* in this regard.

One way to restate the discussion thus far is to consider psychological events (for instance, "functions", "behaviors") and physiological events (for instance, brain regions), which can be denoted $\Psi$ and $\varphi$, respectively [9]. To understand how these two domains are related to each other, one is interested in both $P(\Psi|\varphi)$, that is, the probability of a psychological event given the involvement of some neural structure, and in $P(\varphi|\Psi)$, that is, the probability of a neural event given a psychological one. In other words, what is the mapping between the psychological and physiological domains? In the ideal situation, both $P(\Psi|\varphi) = 1$ and $P(\varphi|\Psi) = 1$, that is, knowledge of the psychological perfectly predicts the physiological and knowledge of the physiological perfectly predicts the psychological. In general, these two probabilities can differ dramatically (they are directly related to each other via Bayes' rule, of course).

The casting of the problem in the above terms is pertinent given the fast accumulation of neuroimaging studies in the past two decades, which are now available to investigators in various databases. For example, Poldrack ([10]; see also [11]) evaluated the common practice in neuroimaging research of drawing *reverse inferences,* namely reasoning backwards from the presence of brain activation to the engagement of a particular function (for instance, "if the amygdala was active, an emotion was involved"). The specific example considered by Poldrack assessed the ability to use activation in Broca's area (in the left ventral-lateral PFC) to predict the engagement of "language function". Via the application of Bayes' rule, which allows one to update prior beliefs based on new evidence, P(Language|Activation in Broca's area) was .69 (based on activations available in the BrainMap database at the time). Therefore, relatively weak evidence (but better than 50/50) was available that given activation in Broca's area, language was involved. Note that Broca's area and language were chosen because, if anything, they would be more favorable to the possibility of reverse inference – given the historical link of this region with language [1, 12]. However, even in such cases, $P(\Psi|\varphi)$ is nowhere near one (the one-to-one case). See Section 11 below for related discussion.

---

[1] See Pessoa [4] for examples concerning emotion and cognition; Cisek [5] for examples concerning perception and action; and Schultz [6] for discussion related to dopamine function.

## 2. Degree of isolability and decomposable systems

To understand the relationship between structure and function, it is instructive to consider architectural features that constrain the mapping between the two. At one extreme, one may consider the concept of a *module* instantiated by a single brain region. In this case, the individual region is solely (or mostly) responsible for carrying out a certain function. One such case is the proposal that the fusiform gyrus in ventral visual cortex of the right hemisphere acts as a "face module" [13]. Here, I follow the discussion by Shallice ( Chapter 11in [1]) and discuss other arrangements that inform the structure-function relationship.

**Processing space continuum.** An example of this architecture can be illustrated by the organization of some sensory regions. For example, consider the retinotopic organization of early visual cortex, namely, the orderly "map" of visual space, where as one moves along the cortex, cells respond to parts of the visual field that move accordingly in external visual space; or the tonotopic organization in early auditory cortex, the orderly map of sound frequency, where as one moves along the cortex, cells respond to different temporal frequencies of an auditory stimulus in a regular manner (say, from low to high frequency). A more interesting example is the hypothesis that the ventral visual system can be conceptualized not in terms of perceptual and memory processes (in posterior and anterior regions, respectively), but in terms of a continuum between the two; in this case, two spatial gradients would exist, such that more posterior brain regions would be more involved in perception (and less in memory), while more anterior brain regions would be more involved in memory (and less in perception). Given this type of organization, patterns of impairment in perceptual and memory functions following brain damage reflect the *demand* that each task places (at specific brain locations) rather than dissociable cognitive modules [14, 15].

**Overlapping processing systems.** Consider a process $P_A$ that requires regions $R_1$ and $R_C$, whereas process $P_B$ requires regions $R_2$ and $R_C$. In this case, $R_1$ and $R_2$ are not parts of two isolable subsystems because they operate as parts of overlapping systems (the overlap involving common region, $R_C$).

**Coupled systems.** When systems are coupled, the *degree of isolability* will depend on the strength of the connections between the two subsystems, which determine how strongly they interact with each other. If they are only weakly coupled, the operations they carry out can be established without considering the behavior of the other system, at least in some contexts. In cases in which the connections between the two are stronger, the coupled systems are only *partially* isolable. In the extreme, the parts are *not* isolable. As discussed below, anatomical connections between brain elements do not need to be strong in order for them to influence one another in important ways. Thus, the *functional connectivity* between a pair of regions, which can be described via the correlation between their respective signals (see Section 7), can be strong even when the structural connection is not.

More broadly, the degree of isolability can be linked to the notion of *decomposable*, *nearly decomposable*, and *nondecomposable* systems [16, 17]. On the one hand, a decomposable system is one in which each subsystem operates according to its own intrinsic principles, independently of the others – that is, it is highly modular. On the other hand, a nondecomposable system is one in which the connectivity and inter-relatedness of the components is such that they are no longer clearly separable. In between, one finds a continuum of possible organizations.

## 3. From brain regions to networks

The previous section considered architectural features that inform the understanding of structure-function mapping from a conceptual point of view. Here, I will briefly review work demonstrating the importance of networks of brain regions. In the next section, I will build from these proposals and describe a structure-function framework. The objective of the current section is not to provide a comprehensive historical account of the notion of networks, but to simply provide a few illustrations. Important treatments not covered here include those by Edelman [18], Grossberg [19], Damasio [20], and Young et al. [21], among many others. For a discussion of the modular viewpoint, please see the treatment by Kanwisher [22], who suggested that "the possibility is within reach of obtaining a cognitively precise parts list for the human brain" ([22]; p. 11168), namely, the list *L* described above. I take a different path here, one that emphasizes coalitions of regions that *jointly* contribute to behavior.

Traditionally, neuroanatomy focused on describing the substrate for the elaboration of neural signals as they progress from sensory through associational to motor centers. For example, in a seminal monkey anatomical tracing study, Jones and Powell [23] studied the convergence of sensory pathways through the analysis of "the sequence of association connections passing outwards from the primary sensory areas as though following the successive steps in a (supposed) sequence of cortical function…to identify regions of convergence within the cortex" ([23], p. 794). This *hierarchical* scheme of cortical organization emphasized the convergence of information leading to integration in sites such as the posterior parietal cortex, superior temporal polysensory area, and prefrontal cortex. As summarized by Goldman-Rakic in an influential paper:

> The conclusion traditionally reached in virtually all comprehensive studies of cortical connections is that they are organized in a step-wise hierarchical sequence proceeding from relatively raw sensory input at the primary sensory cortices through successive stages of intramodality elaboration allowing progressively more complex discriminations of the features of a particular stimulus. ([24] p. 146)

In a considerable departure from this scheme, Goldman-Rakic [24] emphasized, instead, the existence of *distributed* processes carried out via several *parallel* systems. In this scheme, integrative functions emerge from the dynamics of the entire network rather than from computations performed at each nodal point in the circuit.

In an influential series of papers spanning several decades, Mesulam has advanced a network approach to understanding the localization of complex functions as "an alternative to more extreme approaches, some of which stress an exclusive concentration of function within individual centers in the brain and others which advocate a more uniform (equipotential or holistic) distribution" ([25]; p. 309). Mesulam suggested that such network approach would help reconcile some of the inherent problems with the "extreme approaches". In early work [25], he suggested that a network involving the posterior parietal cortex, frontal cortex, and cingulate cortex, contribute sensory, motor, and motivational representations to attentional processes, respectively. In addition, reticular structures in the thalamus and brainstem are involved in arousal aspects of attention. In later work [26], he outlined a more comprehensive scheme in which the human brain contains at least five major "core" functional networks: (i) the spatial awareness/attention network anchored in posterior parietal cortex and the frontal eye field of the frontal cortex; (ii) the language network anchored in Wernicke's and Broca's areas; (iii) the explicit memory/emotion network anchored in the hippocampal–entorhinal complex and the amygdala; (iv) a face-object recognition network anchored in mid- and anterior temporal cortices; and (v) a working memory/executive function network anchored in lateral PFC (and possibly inferior parietal cortex).

The importance of networks also was emphasized by Barbas in the domain of emotion and cognition [27]. She described several anatomical features of the prefrontal cortex that potentially underlie cognitive-emotional interactions. In particular, she proposed that pathways between the amygdala and both orbital and medial PFC provide a means for sensory signals reaching the prefrontal cortex to be integrated with emotional information.

3.1. *Large-scale analysis of anatomical connectivity*

The systematic compilation of anatomical data has revealed massive connectivity between cortical areas [21, 28], between subcortical and cortical areas [21], and between subcortical areas [29, 30]. For instance, Swanson and colleagues reported that on the order of 600 connections of the amygdala were known at the time of their publication [30], and estimated that the total would likely be closer to 1000. More recent work has quantified anatomical connectivity in important ways. Particular interest has been focused on characterizing these data and relating them to specific network topologies. Notably, so-called *small-world* networks [31], which are ubiquitous in natural, social, and technological systems, combine densely clustered connectivity with a small admixture of "random" connections, including *long-range* ones. They preserve a high degree of connectivity within local neighborhoods while allowing *all* nodes of the network to be linked by surprisingly short paths (that is, connection steps), thus creating a "small world" within the network – as in the famous "six degrees of separation" [32]. Several studies have suggested that the cortical connectivity of the macaque exhibits small-world properties [33]. Therefore, pairs of regions are linked by short paths despite large network size and *sparse* overall connectivity (but see further discussion below) – where sparsity refers to the low density of anatomical connectivity.

Recent large-scale network analysis of as many as 383 brain regions (Figure 2A) and cortico-cortical, cortico-subcortical, and subcortico-subcortical connectivity has supported the "small-world" nature of brain connectivity [34]. This work described a tightly integrated "core circuit" (Figure 2B), spanning parts of premotor cortex, temporal cortex, parietal cortex, prefrontal cortex, thalamus, "basal brain" (subcortical nuclei at the base of the forebrain, including the amygdala and basal ganglia), cingulate cortex, insula, and visual cortex. The core circuit was proposed to be "topologically central" – that is, strongly connected to all other regions of the core *and* the rest of the brain – and to have several important properties: (i) it is a subnetwork that is far more tightly integrated than the overall network; (ii) information likely spreads more swiftly within the core than through the overall network; and (iii) the overall brain network communicates with itself mainly through the core. A related concept is that of a *rich-club*, namely a dominant cluster of highly influential nodes [35]. Rich-clubs are relevant because they may determine several important properties of *entire* networks (e.g., the entire brain network). Based on structural human brain data, van den Heuvel and Sporns[36] (p. 15784) proposed that "the aggregation of hubs [i.e., regions of high connectivity] into a rich club suggests that the communication hubs of the brain do not operate as individual entities, but instead act as a strongly interlinked collective" (see also [37]).

--- Figure 2 ---

Several large-scale analyses of anatomical connectivity indicate that the prefrontal cortex contains a disproportionate share of topologically central regions in the brain, as evaluated in terms of measures of efficiency in aggregation and distribution of information, among others. For example, Averbeck and Seo [38] subdivided the prefrontal cortex into 25 areas and characterized how they are connected to each other, in addition to 68 other brain regions. Prefrontal regions exhibit a very high degree of interconnectivity, showing that input information from sensory, motor, and "limbic" (as labeled by the authors) areas can reach anywhere within frontal network within at most two connections.

3.2. *Is brain architecture really small world?*

In a small-work architecture, pairs of elements are linked by short paths despite large network size and *sparse* overall connectivity. If one thinks about it, this is a remarkable property; connections are costly, but short paths are desirable so that elements can effectively communicate with each other – small-worlds have one without the other. In terms of the brain, a small-world architecture would imply that the average brain region would be able to influence most other brain regions via a few connection steps, much as in the PFC analysis of Averbeck and Seo [38]. This property is the more remarkable the sparser the connectivity. But how sparse is brain connectivity?

In a recent study in humans by Hermundstad and collaborators [39] used diffusion MRI to estimate brain connectivity. They found that, of the possible pairings between 600 regions, less than 2% were estimated to be anatomically linked within a given subject, whereas even fewer were consistently linked across subjects. Unfortunately, at the present time, diffusion MRI does not provide precise enough estimates of structural connectivity, and it is hard to interpret estimates of brainwise connectivity. Instead, in a Herculean effort to map cortical connectivity at a more global level, Markov, Kennedy and colleagues injected a retrograde tracer into 29 cortical areas distributed relatively evenly across the macaque cortex [40], allowing them to determine connectivity for a complete 29 x 29 connectivity graph. Though this is subset of the entire 91 x 91 space that they want to map out eventually, their findings thus far paint a very different picture of cortical connectivity, one in which density is far from sparse – they found density to be 66%, namely, 66% of the connections that may exist do in fact exist. Earlier work in macaques also indicates that density is far from low. For example, in the classical analysis by Felleman and Van Essen (1991), connectivity density for visual areas was 32% though they predicted a density of 45% if the unknown connections were to be tested. A few years later, Jouve and colleagues [41] updated the data set analyzed by Felleman and Van Essen and revised those figures to 37% (observed) and 58% (predicted).

Thus, brain architecture seems to be too dense to be small-world. Yet, an important ingredient of small-world organization -- the existence of *non-local* connections, especially *long-range* ones – is present in the brain, too. Although they appear to be relatively weak, long-range connections play a major role in the cortical network. Kennedy and colleagues suggest that they help in communicating "global signals" to small, specific groups of areas and are thus part of a focused "integration mechanism". Notably, in their data, consideration of long-distance connections markedly increased the number of cortical areas projecting to target regions (e.g., extra-visual areas projecting to visual cortex). And contrary to local connections, long-distance ones are specific and thus help determine the "connectivity profile" of their target areas. In other words, long-distance connections are not a random, but instead have specificity in their connectivity targets. Overall, however, connectivity density is concentrated at short distances; long-distance connections are considerably sparser. But, notably, unlike the findings based on MRI diffusion [39], long-distance connections were highly consistent across brains.

In recent computational analyses, Markov and colleagues [42] suggest the intriguing possibility that, at higher densities, rewiring (i.e., randomly inserting/deleting connections) has minimal impact on path length between regions. At higher densities, it appears that density per se determines path length, largely independently of the detailed structure of network connections. According to their simulations, small-world-like properties (reduced average path length while exhibiting high clustering) appear to change at connection densities around 40%.

## 4. Brain architecture viewed from a network perspective

Structure-function relationships can be conceptualized within a network approach (Figure 3). Networks of brain regions collectively support behaviors. Thus, *the network itself is the unit*, not the brain region. Processes *P* that support behavior are not implemented by an individual area, but rather by the interaction of multiple areas, which are dynamically recruited into multi-region assemblies. For instance, Dosenbach and colleagues [43] have proposed that goal-directed executive control can be understood in terms of two networks, a cingulate-operculum[2] network responsible for "set maintenance" (such as maintaining task focus over relatively extended periods of time) and a frontal-parietal network responsible for rapid adaptive control (such as switching between tasks). As another example, Menon, Uddin, and colleagues suggest that a salience network including the anterior insula and the anterior cingulate cortex is involved in attention to external and internal worlds [44, 45]. Naturally, these examples are arbitrary and a growing number of networks are being ascribed particular functions.

---

[2] The operculum is the part of the cerebral cortex that covers the cortex within the lateral sulcus (also called Sylvian fissure), which includes the insula.

--- Figure 3 ---

Networks commonly are described in terms of unique, *non*-overlapping sets of brain regions. But this assumes that brain areas compute a specific function, one that is perhaps elementary and needs other regions to be "actualized", but nonetheless is well defined. The framework advanced here proposes that networks contain overlapping regions, such that specific areas will belong to several intersecting networks [46]. In this manner, the processes carried out by an area will depend on its network affiliation at a given time. What determines a region's affiliation? Here, the importance of the *context* within which a brain region is operating must be considered [47]. For example, in Figure 3B, region $A_N$ will be part of network $N_1$ during a certain context $C_K$, but will be part of network $N_2$ during another context $C_L$. The existence of context-dependent, overlapping networks also means that from the perspective of structure-function mappings summarized in Figure 3B, a given region will participate in multiple processes. I return to the issue of overlap in a later section.

The importance of context emphasizes the need to consider *dynamic* aspects of structure-function relationships. A network needs to be understood in terms of the interactions between multiple brain regions as they *unfold* temporally. In the extreme, two networks may involve the exact same regions interacting with each other in distinct ways across time (Figure 4). Put differently, what matters is the profile of spatiotemporal activity. Consequently, the structure-function mapping is not a static property, but a dynamic one [48]. More broadly, network affiliations evolve across time, such that structure-function mappings are dynamic properties of the brain, in fact taking place over several temporal scales – from hundreds of milliseconds to minutes to days [48-51]. Accordingly, how regions are affiliated with networks and hence the way they impact the behavioral landscape should be viewed as dynamic (Figure 4).

--- Figure 4 ---

Though simple, this point is sufficiently important to merit a few examples. Consider the case of the amygdala. Even a simplified view of its anatomical connectivity shows that, minimally, it belongs to three networks. The first is a "visual network", as the amygdala receives fibers from anterior parts of temporal cortex. Indeed, cells in the amygdala respond to visual stimuli, including faces, with response latencies a little longer than cells in anterior visual cortex (a region that responds to abstract visual stimuli, such as specific shapes). The amygdala, by its turn, influences visual processing via a set of projections that reach most of ventral occipito-temporal cortex. The second is the well-known "autonomic network", as evidenced by connectivity with subcortical structures such as the hypothalamus and periaqueductal gray (these regions are engaged, among others, in the generation of "bodily states", such as those associated with emotional states). Via this network, the amygdala participates in the coordination of many complex autonomic mechanisms (for example, enhancing bodily arousal). The third is a "value network", as evidenced by its connectivity with orbitofrontal cortex and medial PFC. The orbitofrontal cortex, for example, is important for determining the relative value of the current state, and orbitofrontal responses differentiate events, even those removed from actual reward delivery, if they provide information about the likelihood of a future reward. In total, the amygdala affiliates with different sets of regions ("networks") in a highly flexible and context-dependent manner. Many other examples of this *dynamic affiliation* idea exist, including the fronto-parietal cortex, whose regions affiliate with others based on current needs [52].

Two issues deserve further consideration here. First, when describing networks in Figure 4, the term "process" is preferable to "function". One reason is that a *process* is suggested to emerge from the interactions between regions – it is thus an *emergent property* (see [44]). Furthermore, a process is viewed as a useful external description of the operation of the network, and not necessarily as a fixed internal computation implemented by the network [53-55].

A second – and critical -- issue is whether utilizing networks solves the many-to-many mapping problem that is faced when considering regions as the unit of interest. In other words, does a description of structure-function relationships in terms of networks allow for a one-to-one mapping? For instance, as stated above, Menon, Uddin, and colleagues suggested that a salience network involving the anterior insula and the anterior cingulate cortex "mediates attention to the external and internal worlds" ([44], p. 285). They note, however, that "to determine whether this network indeed specifically performs this function will require testing and validation of a sequence of putative network mechanisms…" (p. 285) (see also [56]). The prospect of simpler structure-function relationships (hence *less* context dependent) is discussed by Buckner and colleagues ([57], pp. 1867-8; italics added) when describing regions of high connectivity, at times called "hubs": "An alternative possibility is that the hubs reflect a stable property of cortical architecture that arises because of monosynaptic and polysynaptic connectivity. Within this alternative possibility, the same hubs would be expected to be *present all of the time*, independent of task state."

I suggest that the attempt to map structure to function in a one-to-one manner in terms of networks will be fraught with similar difficulties as the one based on brain regions (Figure 1) – the problem is simply passed along to a higher level. Thus, two distinct networks may generate similar behavioral profiles (Figure 3D; many-to-one); a given network will also participate in several behaviors (one-to-many). Broadly speaking, a network's operation will depend on several more global variables, namely an extended context that includes the state of several "neurotransmitter systems", arousal, slow wave potentials, etc. In other words, a network that is solely defined as a "collection of regions" is insufficient to eliminate the one-to-many problem. What if we extend the concept of a network with these additional variables? For example, Cacioppo and Tassinary [58] suggest that psychological events can be mapped to physiological ones in a more regular manner by considering a spatiotemporal pattern of physiological events – in the notation of the first section, the latter can be denoted as $\varphi_{x,t}$ to suggest changes in space and time. The notion of a network is thus extended to incorporate other physiological events, for instance, the state of a given neurotransmitter (as in the beautiful work by Marder and colleagues; see [59]). How extensive does this state need to be? Clearly, the usefulness of this strategy in reducing the difficulties entailed by many-to-many mappings will depend on how broad the context must be [55].

## 5. Describing and characterizing networks with graph theory

The brain is clearly not an equipotential mesh in which all regions play the same roles [60], so unraveling the contributions of a given brain region to behavior will always be of interest. To understand the impact of a region on behavior, its connectivity pattern should be considered [61]. Intuitively, the extent of anatomical connectivity will be a key element in determining the influence that a region has on brain processing. Accordingly, a region that connects to just a few others will have much less of an impact than one that is more richly connected (other things held constant). The topology of the connectivity will be essential, too. A region with local connectivity will contribute to local computations, whereas a region with more widespread connectivity will have a broader effect.

These considerations can be formalized via graph theoretical concepts [62]. If brain regions are equated with *nodes* and information on structural connectivity is captured by the *edges* between nodes, one can define a node's *degree* by the number of connections of that region[3]. Regions characterized by a high degree of connectivity (say, one standard deviation above the mean of the graph in question), often called *hubs*, are important in regulating the flow and integration of information between regions (area $A_n$ in Figure 3B; Figure 5) [63, 64] ("importance" will be further refined in a later section).

---

[3] This definition applies to so-called binary networks in which edges are either present or not. Degree can also be defined in weighted networks, where edge strength varies continuously.

--- Figure 5 ---

While the number of connections is important in determining whether a region will operate as a hub, its structural *topology* is relevant, too. Some regions are best characterized as "provincial" hubs because they occupy a central position within a single functional cluster [63]. For example, visual area V4 is highly connected with other visual regions, and can thus be considered to be a visual hub [65] – a type of provincial hub. Other regions may be better characterized as "connector" hubs, as they link separate regional clusters. For example, Brodmann's area 46 in dorsal-lateral PFC is extensively connected with other regions, suggesting that it is an important hub in the brain [34, 65] (Figure 6). Because area 46 is highly connected to other prefrontal regions and it is additionally connected to visual area V4, it may operate as a connector hub linking prefrontal and visual regions (see below for further discussion regarding hub-like functions of the lateral PFC). Interestingly, it is also possible to conceptualize area V4 as a connector hub linking visual and prefrontal regions (see also [34]). This last point highlights a key concept: there is never a single way to "slice" networks, a theme to which I will return later.

--- Figure 6 ---

These considerations help understand the repercussions of brain lesions on behavior. The topological characterization of brain connectivity suggests that the impact of lesions will be strongly dependent on a region's structural embedding: lesions of more peripheral (non-hub) regions will produce relatively specific deficits, whereas lesions of hub regions will have a much greater impact on behavior, one that will be strongly determined by the precise topology of the hub (for example, provincial versus connector). In particular, lesions of connector-hub regions will have widespread, and difficult to characterize, effects on cognitive and affective behaviors.

More broadly, a node in a graph can be characterized by a growing number of metrics, several of which attempt to summarize its contribution to information integration and/or distribution. For detailed treatments, please see [62]; for applications to brain data, please see [66]).

## 6. Cortical myopia: The problem of a cortico-centric view of brain architecture

Large-scale analyses and descriptions of brain architecture suggest principles of organization that become apparent only when information is combined across many individual studies. Unfortunately, most of these "meta" studies are cortico-centric – they pay little or no attention to subcortical connectivity. But subcortex is not a simple "outflow" of cortex. Indeed, neuroanatomists have delineated complex cortical-subcortical circuits that have major implications for the understanding of the brain's overall organization. Perhaps the most notable example is that of cortico-basal ganglia systems connectivity to comprise at least five circuits [67]. A fundamental property of these circuits is that they include a "closed loop" component; specifically, the cortical area that projects to the basal ganglia is the same that receives "returning" connections (via the thalamus). More generally, the consideration of subcortical connectivity dramatically alters the computational landscape of the brain, as illustrated next.

### 6.1 Pulvinar nucleus of the thalamus

The pulvinar complex, as this set of related nuclei is sometimes called, is the largest nuclear mass in the primate thalamus and thought to have expanded in size as it evolved in primates[68, 69]. The pulvinar exhibits extensive *bidirectional* connectivity with cortex. For example, all twenty to thirty known visual cortical areas connect with the pulvinar, sometimes in a relatively topographic fashion [70, 71]. Parietal, frontal, orbital, cingulate, and insular cortex are all connected with the pulvinar, too. Remarkably, at a gross level, it is as if the entire convoluted cortex were "shrink-wrapped" around the pulvinar [70]. The dorsal pulvinar has connections with cross-modal association cortex, including temporal areas, and parietal areas that participate in attention. It also receives highly processed visual input from anterior parts of ventral visual cortex. And because it is connected with cingulate cortex, frontal cortex (including orbitofrontal cortex), insula, and amygdala, it has remarkable potential to integrate information from brain regions that are very diverse. Indeed, sites in the dorsal pulvinar may be connected with relatively distal parts of the brain, such as parietal and frontal cortex [70]. Notably, many extensive fronto-parietal cortical connections are mirrored by overlapping fields in the pulvinar [72, 73]: where regions in frontal and parietal cortex are interconnected in the cortex, their projection sites in the pulvinar typically coincide (and the connections are bidirectional between pulvinar and cortex)—an organization that further exemplifies the integration ability of the pulvinar.

Despite considerable progress, understanding of pulvinar function is largely incomplete. One suggestion [74] is that the pulvinar participates in *cortical communication*, with direct cortical connections between two areas supplemented by an indirect pathway coursing through the pulvinar or other higher-order thalamic nuclei (Figure 7). Data consistent with this proposal were reported in rat somatosensory cortex, where activity was found to be driven by a cortico-thalamo-cortical pathway [75]. Results of a monkey physiology study lend further support to the proposal. By recording simultaneously in the pulvinar and cortical visual areas V4 and TEO (the latter in inferior temporal cortex), Saalmann and colleagues [76] reported that maintaining attention in the absence of visual stimulation (during a delay period in which the monkey was preparing for task execution) depended on *pulvino*-cortical interactions. In contrast, direct *cortico*-cortical influences during this delay period were weak (though strong when the cue was shown). It is particularly intriguing that the *relative* contribution of the pulvinar on cortico-cortical interactions was largest during the delay interval. At this juncture in the trial, cortical signals would presumably benefit the most from the support of the pulvinar. In all, as summarized by Theyel et al. [75], "corticothalamocortical information transfer may represent an important addition to, or even replacement of, the current dogma that corticocortical transfer of primary information exclusively involves direct corticocortical pathways."

--- Figure 7 ---

## 6.2. Amygdala

The primate amygdala is massively interconnected with cortex. Based on the available data at the time, analysis of amygdala pathways by Young and colleagues [21] revealed that, of the 72 cortical areas included in their study, it was connected to all but eight (see also [27, 77]). Indeed, Petrovich and colleagues [30] estimated that the amygdala may have as many as 1,000 separate cortical and subcortical pathways. The connectivity is all the more notable given that it involves all cortical lobes as well as subcortex. Combined, these properties indicate that the amygdala is an extensively interconnected *hub* region. Furthermore, in the network analysis by Modha and Singh [34] discussed previously, several amygdala nuclei (e.g., lateral nucleus, accessory basal nucleus) were identified as part of a *core* brain circuit, all of whose regions have extremely high connectivity, indicating that the amygdala is part of a "rich club" of regions. Together, these findings reveal that the amygdala has exceptional potential for information integration and distribution.

Consider an example of what the connectivity may accomplish functionally: the amygdala is capable of integrating diverse signals to help *determine* value (as briefly discussed earlier) – such as the biological importance of stimuli. The amygdala receives highly processed sensory information [21, 78], which is important for value representation since it often depends not on superficial sensory properties, but on those showing some "invariance." For example, a dominant male's face is biologically significant irrespective of viewing angle, illumination, and the like. Also relevant, the amygdala receives signals from all sensory modalities [79]. Thus, for example, the significance of a dominant male's vocalizations should be evident by recognizing that they were uttered by him and not by another male further down the hierarchy. In all, the amygdala participates in a large number of cortical-subcortical circuits that are important for the determination of value.

Another property of amygdala connectivity is to support the *broadcasting* of value-related signals across the brain. In this regard, the pattern of connectivity between the amygdala and prefrontal cortex [80] is of particular interest given the latter's role in cognitive functions. In addition to substantial connections between the amygdala and both medial and orbital aspects of prefrontal cortex, recent findings indicate that *bidirectional* connectivity is present also along the lateral surface [81]—although the connections are relatively weak. More generally, to understand how amygdala signals are potentially communicated to all sectors of prefrontal cortex, one must consider PFC connectivity itself. In one study, although the amygdala was estimated to be directly connected to approximately 40 percent of prefrontal regions, approximately 90 percent of prefrontal cortex was shown capable of receiving amygdala signals after a single additional connection *within* prefrontal cortex (Figure 8, [38]). This "one-step" property seriously undermines the notion that amygdala signals are confined to orbital and medial PFC territories, as often assumed in the literature. And arguments that amygdala outputs have limited influence on responses in lateral PFC cortex have less traction once we consider these architectural features. Of relevance, in a functional MRI study by Messinger and colleagues, microstimulation of the monkey amygdala generated responses in lateral PFC, including dorsal sites [82]; thus, under normal brain functioning, amygdala signals may impact these brain locations. Although the downstream effects of microstimulation may involve both mono- and polysynaptic connections, this finding demonstrates that the amygdala has the ability to influence lateral PFC (a region that is frequently conceptualized as "cognitive").

--- Figure 8 ---

In summary, the connectivity pattern between the amygdala and prefrontal cortex reveals ample opportunities for signal communications (Figure 8); and, through couplings of varying strengths, amygdala–prefrontal cortex interactions enable cognitive-emotional interactions and integration [2]. Other notable aspects of amygdala connectivity not discussed here include interactions between the amygdala and the basal forebrain that are important for attentional functions and substantial projections from the amygdala to visual cortex that influence competition in visual cortex, thus contributing to attention.

### 6.3. Hypothalamus

The hypothalamus is involved in several important "basic" operations. For example, it coordinates many complex homeostatic mechanisms, such as hormonal and behavioral circadian rhythms, in addition to neuroendocrine outputs. Historically, the hypothalamus has been conceptualized in terms of "descending" systems (that is, connecting to regions "downward" from it along the brainstem), a view that is summarized by its designation as the "head ganglion" of the autonomic nervous system. However important the hypothalamus may be for descending control, a significant recent insight is that mammalian cerebral cortex and the hypothalamus share massive *bidirectional* connections. In the rat, the best-studied mammal species, there are four major routes from the hypothalamus "up" to cerebral cortex (Figure 9,[29]). These involve a robust direct projection to *all* parts of the cortical mantle, and indirect routes by way of the thalamus, basal nuclei (specifically, basal forebrain and amygdala), and brainstem.

--- Figure 9 ---

The hypothalamus is probably the largest source of nonthalamic direct input to cortex, as reviewed in detail by Larry Swanson and colleagues [29, 83]. In the rat, some notable targets of hypothalamic fibers include regions in medial PFC and insular cortex. Interestingly, less prominent projections of the hypothalamus to lateral PFC and even to primary sensory areas have been found. The connections between prefrontal cortex and the hypothalamus have also been investigated in nonhuman primates [84], where they were found to closely resemble those observed in rats. And, as stated by Rempel-Clower and Barbas, "the hypothalamus has widespread projections to *all* sectors of the prefrontal cortex. Retrogradely labeled neurons in the hypothalamus were found for every tracer injection on the orbital, medial, and *lateral* surfaces"(p.413; italics added[84]) .

Therefore, whereas the hypothalamus is involved in a host of basic control functions, it is part of an extensive bidirectional connective system with cortex and many other subcortical structures in a manner that allows for integration of wide-ranging signals. Critically, the hypothalamus is linked to other structures that have themselves broad connectivity, including the basal forebrain (see [2]) and the amygdala, further expanding its potential for influencing information processing.

The examples reviewed here – pulvinar, amygdala, and hypothalamus – illustrate how subcortical structures interact with cortex in ways that cannot be neglected when attempting to discern principles of brain organization. The *cortical*-based analysis of brain connectivity typical of many studies provides a description that is not only incomplete but, worse, highly distorted.

### 6.4. Evolution and cortical-subcortical co-embedding

To further understand the contributions of subcortical connectivity, it is worth briefly considering the evolution of the brain. A traditional view is that cortex is a late addition to the brain plan and that it controls subcortex. Indeed, the idea of cortical inhibition of subcortex has a long history dating to early researchers, such as Hughlings-Jackson (see [85]).

But what is the *basic plan* of the vertebrate brain? It is now understood that both cortex and subcortex are part of the plan. The brain can be described in multiple ways, but Figure 10 shows a proposed brain "archetype" by Striedter [86]. The main features are common to all vertebrates. Another way is via a "flat map" as proposed by Swanson (Figure 11). In this case, the brain's overall plan is captured in nine elements (plus the spinal cord, also shown). Thus, the plan of the vertebrate brain is the same for *all* vertebrates – brain evolution is a *conservative* process [87].

--- Figures 10 and 11 ---

A cortico-centric framework is one in which the "newer" cortex controls subcortical regions, which are typically assumed to be relatively unchanged throughout evolution. In this view, cortical expansion is thus a matter of cortical regions being set up so as to control "lower" centers. In sharp contrast, if both cortex and subcortex change, they may change in a coordinated fashion. In this case, the resulting circuitry is one in which cortex and subcortex are mutually *embedded*.

A good example of this type of mutual embedding is the amygdala. The amygdala of mammals is composed of a dozen or more subregions. Chareyron and colleagues [88] found that the lateral, basal, and accessory basal subregions are dramatically more "developed" in monkeys than in rats (based on morphological characteristics, such as cell counts and the volume of subregions). One possibility, as described by the authors, is that the differences in the relative size and neuron numbers between rats and monkeys are linked to their degree of connectivity with other brain structures, in line with the proposal of correlated evolution between components of functional systems [89]. The lateral, basal, and accessory basal nuclei are more developed in primates than in rodents, and parallel the greater development of the cortical areas with which these nuclei are interconnected in primates. Chareyron and colleagues propose that such correlated evolution may be responsible for a higher convergence and integration of information in the primate amygdala, and that the relative development of these amygdala nuclei might be influenced by their interconnections with other brain structures, namely, their afferent and efferent connections [78][4] (Figure 12).

--- Figure 12 ---

Work by Barger and colleagues [90] have shown that the lateral nucleus of the amygdala in humans is significantly larger than predicted for an ape of human brain size. Indeed, the differences observed were quite dramatic, with a magnitude rarely seen in comparative analyses of human brain evolution. Neuron numbers in the human lateral nucleus were nearly 60% greater than predicted. By comparison, the volume of the human neocortex is 24% larger than expected for a primate of our brain size [91], whereas the human frontal lobe, frequently assumed to be enlarged, is approximately the size expected for an ape of human brain size [92].

## 7. Functional connectivity

Thus far, the discussion has emphasized the role of structural connectivity, which provides the physical backbone for functional relationships between regions. At a first glance, the notion of an architecture anchored on physical connections is clear cut. However, the boundary between anatomy and function becomes blurred very quickly once one starts considering factors that characterize the anatomy [93]. For instance, the type of receptor subtypes involved, the laminar profile of the connections (often interpreted in terms of "modulatory" vs. "driving" inputs), the presence of excitatory or inhibitory interneurons (and the ratio of these), the strength of the connection, and so on. The existence of complex circuits, which include multiple feedforward and feedback connections, diffuse projections systems, etc., further complicate the picture.

Understanding how regions and networks contribute to brain function thus requires identifying the way regions are "functionally connected". *Functional connectivity* is a concept that was initially devised to characterize how neurons interact and was defined as the "temporal coherence" among the activity of different neurons, as measured by cross-correlating their spike trains [94-96]. It is defined also as the "temporal correlation between spatially remote neurophysiological events" [97]. Hence, functional connectivity is essentially a "model free" description of the *joint state* of multiple brain elements (for instance, neurons, areas, etc.). Functional connectivity is at times contrasted with *effective connectivity*, which attempts to explain the origins of the observed functional relationship and, by definition, assumes a model [98]. For instance, when applied to neurons, effective connectivity was initially defined as the simplest neuron-like physical circuit that would produce the same temporal relationship as observed experimentally between two neurons in a cell assembly.

The relationship between structural and functional connectivity is a complex one. For instance, in principle, responses in two regions could be perfectly correlated (barring, say, noise) due to common inputs (Figure 13B). They also could be perfectly correlated, yet having the effect be entirely mediated via an intermediate region (Figure 13C). More generally, at least three types of relationship should be considered between structural and functional relationships. First, a structural connection that is paralleled by a functional connection. This is the "default condition" that is frequently assumed in the "reverse" direction as well, namely, a functional relationship between two regions is likely the result of relatively direct structural connectivity. Second, a functional relationship between two regions exists that is *not* directly supported by a structural pathway (Figure 13D). Third, a structural, excitatory connection between two regions could exist in the absence of a detectable functional relationship. Although this situation is less intuitive than the previous ones, many scenarios are conceivable. For instance, in Figure 13D, $R_1$ and $R_2$ are physically connected but whereas $R_1$ is robustly engaged during context $C_1$, $R_2$ is engaged not only during context $C_1$, but also during a range of other contexts, some of which, say, lead to a suppression of baseline levels of activity in the region. Accordingly, unless a study focuses exclusively on context $C_1$, a functional relationship between $R_1$ and $R_2$ will not be readily apparent.

---

[4] Note, however, that parsing the reasons for changes in brain size is not possible at the moment and will always constitute a major challenge. Both "concerted" and "mosaic" evolution proposals have been described [86].

--- Figure 13 ---

A variant of the third scenario was illustrated by He and colleagues [99]. They reported a case in which functional connectivity between regions was disrupted in the absence of anatomical damage to those regions or their connections – specifically, inter-hemispheric functional connectivity of the posterior cortex (intraparietal sulcus) in spatial neglect patients. This shows that anatomical connectivity may be *necessary but not sufficient* for normal functional connectivity; excitatory/inhibitory neuronal inputs from other regions must be considered, too. Another illustration is given by a study of patients with multiple sclerosis, a pathology that compromises central white matter in a diffuse manner [100]. In the patients, functional connectivity increased in the face of a concomitant *reduction* of anatomical connectivity (see also [101]).

Neuroimaging studies reveal an important way in which structure and function deviate (Figure 13E): the functional relationship between two regions is *context dependent.* In this case, the functional link changes without any concomitant modification in structure. For example, the functional connectivity between two regions with known (or presumed) structural connections can increase or decrease as a function of several variables, including task performance [102], motivation [103], and emotion [104]. In one study, the functional connectivity pattern between early visual areas was investigated during affective and neutral contexts [105]. During the affective context, participants viewed faces that were surrounded by a ring whose color signaled the possibility of mild shock (Figure 14A). During the neutral context, faces appeared surrounded by a ring whose color signaled safety. A measure of functional connectivity was strengthened during the affective relative to a neutral context (Figure 14B). Thus, the affective context not only changed the magnitude of evoked responses but also altered the *pattern* of responses across early visual cortex.

--- Figure 14 ---

A growing number of studies is seeking to explore the relationship between functional and structural connectivity [106]. Greicius and colleagues [107] employed both tractography[5] and resting-state functional connectivity[6] to compare the two [107]. Whereas some of their findings supported a direct relationship between structural and functional connectivity, several discrepancies were observed, too. In early studies, robust inter-hemispheric correlation of functional MRI signals was observed between cortical regions that have few direct connections, such as those of the left and right hand representation in primary motor cortex [110] and those of spatial representations in left and right primary visual cortex [111]. Koch and colleagues compared anatomical and functional connectivity between cortical patches on adjacent cortical gyri [112]. Low functional connectivity rarely occurred in combination with high anatomical connectivity. In contrast, high functional and low anatomical connectivity did occur in combination.

---

[5] Tractography refers to computational techniques to estimate major white matter fiber tracts based on diffusion-weighted imaging [108].

[6] Resting-state connectivity refers to methods that evaluate functional connectivity across brain areas during conditions that do not involve overt tasks, that is, during "rest" [109].

A striking example of structure-function dissociation was reported by Tyszka and colleagues [113] who investigated an unusual population of adults without the corpus callosum, which provides the major communication pathway between the two hemispheres. Although this group was very different structurally relative to controls, they exhibited very similar functional networks during the resting state. As summarized by the authors: "The present findings argue that largely normal functional networks can emerge in brains with dramatically altered structural connectivity. … Perhaps the most profound aspect of the present findings is the suggestion that the functional organization of the brain subserving cognition can be driven by factors other than direct structural connectivity." (p. 15161).

A recent lesion study with monkeys helps clarify these findings further [101]. The study compared disconnection between the hemispheres by cutting the corpus callosum with and without sparing of the anterior commissure; the latter is a small group of fibers also connecting the hemispheres. They found that the corpus callosum section disrupted inter-hemispheric functional connectivity if the anterior commissure was also sectioned. However, functional connectivity across the whole brain was essentially preserved when the anterior commissure was spared. These findings provide support for a complex relationship between even relatively minor structural connections and widespread functional connectivity. In this case, although the anterior commissure carries inter-hemispheric structural connections between only a subset of areas (the temporal lobes, the orbitofrontal cortex, and amygdala), inter-hemispheric functional connectivity was maintained between a much wider set of regions, including those with no anterior commissure projections (e.g., parietal cortex and dorsolateral prefrontal cortex).

So what determines functional connectivity if structural connectivity is a weak predictor? Relatively little is known at present, but some recent inroads have been made. For example, Adachi et al. [114] compared existing data on structural connectivity in macaques (data collated in the CoComac database) and functional connectivity obtained during MRI scanning of macaques under anesthesia. They analyzed the effect of different types of indirect structural connections on functional connectivity. Remarkably, functional connectivity between pairs of regions *without* a direct cortico-cortical connection depended more strongly on whether two regions (A and B) had common inputs and outputs (A ← C → B; A → C ← B) than on whether there was stepwise information flow between them (A → C → B) (Figure 15). Adachi and colleagues thus proposed that functional connectivity depends more strongly on network level than on pairwise interactions. See also [115, 116].

--- Figure 15 ---

A final consideration in understanding functional connectivity pertains to the "dynamics" implemented in particular regions. For example, it is well documented that lateral PFC circuits can implement reverberating activity that can be sustained for several seconds [117]. Such dynamics are not only important for extending the repertoire of lateral PFC computations but they also influence the precise form of functional connectivity that the lateral PFC entertains with other regions [49]. Therefore, the elucidation of functional interactions requires not only greater knowledge of structural connectivity properties, but also of how local physiological properties impact both short- and long-range brain interactions. Although very little is known about how local computations impact functional connectivity, this is an important question for future research – and one that further exemplifies the complex relationship between structural and functional connectivity.

# 8. Functional networks

In the same way that structural information is used to investigate brain networks, functional connectivity can be used, too. For example, based on patterns of co-activation across a large number of studies, Toro et al. [118] identified three functional networks: a fronto-parietal attention network, a resting-state network, and a motor network. Using a similar approach, Postuma and Dagher [119] identified patterns of co-activation between the cortex and striatum that appear to reflect the cortico-striatal loops proposed by Alexander and colleagues [67]. In one study of resting-state activity [120], the pattern of co-activations between 90 regions (involving major cortical gyri and subcortical nuclei) were submitted to hierarchical clustering, which identified six major systems corresponding approximately to four cortical lobes, a medial temporal cluster, and a cluster comprised of subcortical nuclei. Interestingly, analysis of *functional* connections between regions was consistent with a small-world topology – there were a number of functional connections between regions that were much stronger than would be predicted as a function of the *anatomical* distance between them.

In the past decade, graph-theoretical analysis of functional neuroimaging data has focused almost exclusively on characterizing the large-scale properties of resting-state data [121, 122]. In a recent study, we sought instead to understand the network properties of a focused set of brain regions during task conditions engaging them [123]. Graph-theoretic network analysis was used to characterize how emotional and motivational stimuli potentially alter functional connectivity. Two separate tasks were investigated in which emotional or motivational cues preceded the execution of a response-conflict task. In the emotion task [124], participants viewed an initial cue that determined whether they were in a threat or safe trial; in the motivation task [103], they viewed an initial cue that indicated whether they were in a reward or control trial.

The effects of emotional and motivational cues exhibited several similarities. At the network level, *global efficiency* (a measure of integration) increased and *decomposability*[7] (a measure of how easily a network can be divided in terms of smaller subnetworks or "communities") decreased. In other words, the network became less segregated with the context signaled by the cue (possible shock in one experiment, possible reward in the other), revealing that one way in which emotional and motivational processing affect brain responses is by increasing functional connections across brain regions.

The dual competition model [125] proposes that the effects of reward during perception and cognition depend in part on interactions between valuation regions and fronto-parietal regions important for attention and executive control. Such interactions lead to the up-regulation of control and improve behavioral performance during challenging task conditions (and higher likelihood of reward). The increased functional connectivity between the two communities detected in the motivation dataset (Figure 16A) is consistent with these ideas, and they also suggest that the increases in connectivity can be quite broad. For example, the caudate (Figure 16B) and the nucleus accumbens showed increases in functional connectivity to nearly all cortical regions that were systematically driven by reward.

--- Figure 16 ---

A related pattern was observed with the emotional manipulation. In this case, we observed enhanced functional integration *between subcortical regions* (such as the bed nucleus of the stria terminalis and thalamus) *and cortical regions* (including the insula and medial PFC). But in the case of reward, functional connectivity *increased within cortex*, whereas in the case of threat, functional connectivity *decreased within cortex* for several pairs of regions (Figure 17). How should we interpret the latter findings? The dual competition model proposes that emotional processing diverts resources that also are needed for executive function. The interference is then proposed to impair cognitive performance. Therefore, reduced functional connectivity among some of the cortical areas may have reflected the interference that the threat of shock exerted on subsequent cognitive performance (the conflict task).

---

[7] The technical term used in network science is *modularity*, which is used in a semantically neutral fashion. But this is a tremendously loaded term in the context of neuroscience, hence a "bad" term.

--- Figure 17 ---

Our findings revealed several ways in which both emotional and motivational processing altered functional connectivity, including increased global efficiency and reduced decomposability. An analysis of MEG data has suggested that greater *cognitive* effort is associated with the emergence of a less modular network topology [126]. Given that MEG was used, it is likely that the changes observed were more directly tied to cortical processing (because the technique is less sensitive to deeper brain signals). The results of our study show that the processing of emotional and motivational stimuli may have a similar impact on network organization but emphasize enhanced vertical, *cortical–subcortical functional integration* in a manner that may be behaviorally appropriate. Potential reward may contribute to improved task performance (and reward attainment), and potential threat may redirect mental resources in the service of mobilizing the body toward safeguarding the organism against harm.

## 9. Understanding "importance": Structural and functional embedding

A network framework of understanding brain organization moves the unit of analysis away from brain regions and distributes it *across* them. This does not mean that regions provide equal contribution to specific behaviors, of course. That being the case, devising ways to characterize a region's *importance* is of great interest. Here, I build upon a recent discussion by Vlachos and colleagues [127], who considered this question in the context of neurons.

Neurons whose responses are systematically related to a task are typically assumed to play a role in the underlying computations. In practice, some form of statistical test is applied to establish the significance of the association. However, Vlachos and colleagues [127] provocatively suggest that it is necessary to go "beyond statistical significance"; it is necessary to consider also the implications of *network structure* to neuronal activity. They illustrate how "not all observed activity modulations of neurons in a task are relevant for the specific task itself". Thus, "statistical significance of recorded neural events is only a *necessary* but not *sufficient* condition for making inferences regarding the functional importance of these events for the computations performed by the investigated brain area" (p. 4).

How should one go about determining the *importance* of a neuron to a given computation? One strategy is to consider their degree of *structural embeddedness* and *functional embeddedness* (the latter was denoted "effective" embeddedness in [127]). The former refers to the way neurons are physically embedded in their surrounding network; the latter is the influence neurons have on the activity of the surrounding network, which depends on structural embeddedness, in addition to other synaptic and cellular properties, ongoing activity, neuromodulators, etc. Thus, the importance of task-related neurons is governed by their relative position in the topological space of the network.

This reasoning can be applied to brain areas and not just individual neurons. Let us discuss the case of structural embeddedness. Regions (that is, nodes) with high connectivity (that is, high degree) have the potential to be influential, particularly if they function like connector hubs, as described earlier. Counter intuitively, however, in some circumstances influence does not correspond to the most highly connected nodes of a network ([128]; see also [129]). Instead, the most prominent nodes are those located within the *core* of the network. In other words, they belong to a *topologically central* subnetwork[8]. Hence, one way to measure embeddedness is to determine nodes that exhibit the property of *centrality* [62] (roughly, they are "central" in that they connect to many other nodes, such as an influential member of a social network).

In general, no single measure will perfectly capture *influence* or *importance* because different measures will convey different aspects of network structure. Indeed, multiple measure of centrality have been proposed [66] and generate different results [131]. Therefore, a combination of different metrics will provide a better measure of embeddedness and, better still, how a node affects network properties [34, 127, 129]. In this context, Power and colleagues [132] discuss the shortcomings of using the measure of *degree* to define node importance and suggest other measures instead. Intriguingly, when node degree is employed, the "task-negative" network (engaged in the absence of externally driven, effortful tasks) appears as a critical "backbone" that is central. When other measures are employed, the "task-positive" network (engaged during externally driven, effortful tasks) is highlighted.

A further issue concerns characterizing *communicability* in complex networks [133]. Many important measures that characterize networks are based on the shortest paths connecting two nodes, including the determination of communities. Counter intuitively, as described by Estrada and Hatano [133], "information" can in fact spread along non-shortest paths (see also [134]; [135]). This has implications for the understanding of brain networks because direct anatomical connectivity is frequently emphasized as the chief mode of communication between brain regions. The notion that communicability does not necessarily rely on shortest paths reminds us of the need to obtain network-level properties in describing the flow of signals in neural networks. It also highlights the need to characterize functional connectivity between regions, which does *not* uniquely depend on direct anatomical connections.

### 9.1 The importance of weak connections

Here, I would like to critique another component of the "standard" view, which can be summarized as follows: network states depend on strong structural connections; conversely, weak connections have a relatively minor impact on brain states.

Schneidman and colleagues [136] recorded simultaneously from 40 cells in the salamander retina. Although some pairs of cells had very strong correlations, most correlations were weak. Because the correlation between most cells was small, Schneidman and colleagues tried to infer the whole "network state" via an approximation in which the cells are independent from one another. Whereas this was a good approximation for many pairs, it completely failed when the approximation was applied to the entire population of cells. To illustrate this, the instantaneous state of the network can be described by a binary word (one bit per cell). The "independence model" makes simple predictions for the rate at which each word should occur. The authors found that, for example, at one extreme, the word "1011001010" for a 10-cell set occurred once per minute, whereas the independence model predicted that this should occur once per three years. Conversely, the word "1000000010" was predicted to occur once per three seconds, whereas in fact it occurred only three times in the course of an hour. In conclusion, the independence model makes order-of-magnitude errors (even for very common patterns of activity). Most importantly, the findings of this study demonstrated how *weak pairwise correlations* at times are capable of generating *strongly correlated network states*.

The lesson to be learned here is that weak connections cannot be disregarded when the goal is to understand network states. If these ideas hold more generally, they have important implications for the study of brain architecture because most studies of large-scale networks based on structural and functional data disregard weak connections. In fact, in resting-state functional connectivity studies, researchers typically assign connections with weak correlations (say < .3) a value of zero (that is, no connection) (e.g., [137]). Although, more studies are evidently needed to examine the contributions of weaker connections, their importance is not restricted to cells in the salamander retina. For example, Bassett and colleagues [51] studied the dynamic reconfiguration of human brain networks during learning and uncovered several clusters of brain regions that remained integrated with one another by a complex pattern of weak functional interconnections.

---

[8] One way to evaluate this is to use k-shell decomposition [130]. Intuitively, core decomposition recursively "peels off" the least connected nodes to reveal progressively more closely connected subnetworks.

## 10. Networks: further issues

This section discusses additional issues concerning network analysis. They illustrate that although networks advance our understanding of structure-function relationships in the brain, important questions remain regarding the application of existing techniques. Encouragingly, research on network theory has advanced rapidly since the publication of seminal papers in the late 1990s and new developments are routinely being reported.

To motivate several issues discussed in this section, consider the following statement: "The whole cerebral cortex in fact can be divided into a finite number of RSN [resting-state networks] that maintain relatively stable level of correlation, both within and across networks, along putative axes of functional organization" (p. 782, [138]). This viewpoint – discrete, stable networks in general – is the dominant one in the field.

### 10.1 Network partitioning (community assignment)

In network science, a key question involves the determination of densely interconnected groups of elements, known as *communities*[9]. For instance, Figure 2 displays a decomposition of brain regions in terms of frontal, parietal, cingulate, etc., communities. Most partitioning schemes parse individual elements (brain regions in a brain network, persons in a social network, etc.) into unique communities – for example, Brodmann's area 46 belongs to the "frontal community". Although prevalent, in general, this approach is inadequate given that in many complex systems, elements naturally belong to multiple communities [139, 140]. This is the case in social networks, for instance, where individuals may be a member of several different communities simultaneously, each of which is characterized by a different type of relationship (family ties, work connection, favorite sports team, etc.).

Communities are determined by subdividing a network into groups of nodes, often by maximizing the number of *within*-group links, and minimizing the number of *between*-group links [141]. In other words, it is typically accepted that a community should have more internal than external connections, a property that resonates with the notion of a dense group of elements (brain regions, people, etc.). But consider, for instance, the communities around the word "Newton" in a network of commonly associated English words [140]. "Newton" is part of many intersecting communities. Critically, allowing "Newton" to participate in several communities provides a *better* description of its relationship to other words than forcing it to belong to a single one. Thus, the multiple relationships of "Newton" are simultaneously captured, instead of focusing on a (possibly) dominant one – say, one for which "Newton" has the greatest strength of membership. This means that, counter intuitively, highly overlapping communities can have more external than internal connections [140]. In fact, overlapping communities lead to dense networks and prevent the clear-cut determination of the attribution of individual nodes – that is, the assignment of the community to which a node belongs[10].

As in many applications of network theory, the investigation of brain networks has focused on establishing unique partitions – that is, *non-* overlapping communities – of regions. For instance, in the work of Modha and Singh [34], regions in the frontal lobe comprised a frontal community, regions in parietal lobe comprised a parietal community, and so on. Earlier work that did not rely on algorithmic methods of parsing brain regions into subnetworks, also favored describing non-overlapping groupings. For example, Mesulam [26] suggested that at least five major networks can be identified in the human brain. The networks cover a large extent of brain territory, but are supposed to be relatively separable[11]. In another example, Bressler and Menon [44] described three networks, a "default mode" (resting-state) network, a salience network, and a central-executive network, that are non-overlapping. Whereas these are valuable proposals of how multi-region coalitions carry out mental functions, it is unclear whether non-overlapping decompositions reflect the best way to group cortical and subcortical regions. Given that a more rigorous characterization of brain networks is in its infancy, it is too early to tell. Nevertheless, I believe that insisting on a non-overlapping partitioning scheme provides a poor characterization of brain organization (see [142, 143]). An illustration of this inadequacy is the network overlap observed at connector hub regions, which are especially important for combining diverse sources of information [37]. Consider, for example, a connector hub region as shown in Figures 5 and 6. Whereas a traditional, non-overlapping decomposition will place this region in one of the two partitions, a better way to describe the architecture is one in which the hub belongs to *both* communities (consider also the example of the amygdala provided previously in which it participates in at least three major networks). Put more generally, the characterization of community overlap and how this depends on context is an exciting direction for future research. Finally, community overlap has vital implications for the understanding of structure-function relationships because region overlap generates several of the difficulties reviewed during the discussion of region-to-function mapping (Figure 1) – overlap by definition compounds the one-to-many problem.

---

[9] Although these are at times also called "modules", the term is relatively neutral and is not to be interpreted as a module as discussed in cognitive science, for instance. See note 7.

[10] More technically, this issue is closely related to the existence of structure at multiple scales simultaneously [140]. See also next section.

---

[11] Though Mesulam [46] also discussed network overlap.

## 10.2 Multi-scale structure

A second issue addressed here is that complex networks contain structure at *multiple scales* simultaneously. Situations like this are frequently encountered in the brain. For example, the "visual network" can be decomposed into ventral and dorsal networks [144]. Also, cortical networks involved in executive control can be decomposed into "stable set control" and "rapid adaptive control" subnetworks, as previously mentioned [145]. Multi-scale information naturally implies that one can group regions into hierarchies. The existence of hierarchical structure means that it is necessary to go beyond establishing simple communities. It is necessary to explicitly account for information at multiple scales *simultaneously*. For example, in the study of Averbeck and Seo [38], not only were prefrontal communities detected, but also some of the structure within them.

Let us now consider a methodological issue that is central to the understanding of community-detection methods in general. In computing network partitions, one chooses the "best" partition among a larger set of candidate subdivisions by optimizing a "quality function" that describes the value ("quality") of assigning each node to a specific community. However, in complex networks, many decompositions exhibit "quality scores" very similar to the "best" candidate partition (see Fig. 9 of Good et al., [146]), indicating that the latter is not inherently superior to other decompositions. As summarized by Good et al. (p. 046106-10, [146]): "There are typically an exponential number of structurally diverse alternative partitions with modularities very close to the optimum (the degeneracy problem)." In other words, the quality function landscape will exhibit plateau-like regions. In a similar vein, most real-world networks have multiple plausible hierarchical representations of roughly equal likelihood [146, 147]. Averbeck and Seo [38] proposed an interesting strategy to tackle this problem. To unravel the organization of the prefrontal cortex, they built a "consensus" prefrontal tree generated from the 50 top most-likely trees. Intriguingly, the consensus tree contained many features not observed in the "best" tree, although several noteworthy properties were commonly found in the top trees. Another approach, in this case in the context of standard non-hierarchical analysis, was described by Hermundstad and colleagues [39], who averaged the values of several network diagnostics across one hundred partitions.

More generally, there is no single inherently "best" solution to the decomposition problem, and multiple representations of the structure of a network convey important information. It is thus better to see the representation as a *family of decompositions*. Thus, treating specific decompositions as inherently superior is unwarranted without justification. In real-world situations, these considerations are considerably more acute because data are noisy and imperfect.

A related issue arises in the context of hierarchical decompositions. Whereas hierarchical decompositions are frequently attempted for brain data, in many the hierarchy may be understood as, in fact, *indeterminate*. This is powerfully illustrated in the analysis of the connectivity pattern of the visual cortex. The cortical lamination pattern between the visual thalamus and area V1 has been used as a guide that is extrapolated to other connections, which are then labeled as "feedforward" or "feedback". The former are those that originate in superficial cortical layers and terminate in layer IV; the latter are those that originate from inferior layers and terminate in layers I and VI (and avoid layer IV). Based on these assumptions, Hilgetag and colleagues [148] investigated the organization of the visual system and concluded that "the information in the anatomical constraints cannot be expressed satisfactorily by any single hierarchical ordering" [148]. In this sense, the visual hierarchy might be considered *indeterminate*, such that "no single hierarchy can represent satisfactorily the number and variety of hierarchical orderings that are implied by the anatomical constraints". I propose that this conclusion applies to many other decompositions of brain data[12].

---

[12] Some researchers have suggested that hierarchical information can be resolved once laminar information is established for the regions involved (thus presumably helping determine "feedforward" and "feedback" connections") [149]. In my view, hierarchical views of brain organization are highly problematic (e.g., Chapter 3 of [2]).

### 10.3 Multi-relational structure

Thus far, we have considered networks that are unirelational in the sense that the connection between two nodes represents one kind of information. However, networks are often multirelational (also called multiplex), like when several types of information define the interactions between two individuals in social networks [150]. In the case of brain data, an analogous situation exists: anatomical connections are characterized by strength, laminar profile (that is, given that cortex is layered, cortical layer of origin and target cortical layer), type of neurotransmitter, etc. In addition, functional connectivity information (as defined previously) may be available, too. Algorithms of community detection for multirelational networks offer exciting new avenues to explore network structure [151].

In this section, I discussed a few issues that must be considered when studying brain networks: network partitioning, multi-scale structure, and multi-relational structure. The upshot is that decomposing brain data into "networks" poses many challenges. In the end, both hierarchical and non-hierarchical algorithms (for instance, overlapping community detection) need to be used (separately and in combination) to advance the understanding of brain networks.

### 11. Understanding a region's function via multidimensional profiles

Let us return to the structure-function problem, again from a region's perspective. If, as advanced here, brain regions are engaged in many processes based on the networks they are affiliated with in particular contexts, they should be engaged by a range of tasks. Although this introduces outstanding problems, the availability of data repositories containing the results of thousands of neuroimaging studies provides novel opportunities for the investigation of human brain function [152].

Recently, we employed a data-driven approach to investigate the functional repertoire of brain regions based on a large set of functional MRI studies [153]. We characterized the function of brain regions in a multidimensional manner via their *functional fingerprint* [61]. Activations were classified in terms of 20 *task domains* chosen to represent a range of mental processes, including perception, action, emotion, and cognition, as developed in the BrainMap database [154]. The functional fingerprint for a given region thus represented both the set of domains that systematically engaged the region and the relative degree of engagement. Functional fingerprints of sample regions are illustrated in Figure 18 (top).

--- Figure 18 ---

Based on fingerprints, we calculated a *diversity index* to summarize the degree of *functional diversity*. A brain region with high diversity would be one engaged by tasks in many domains, whereas a low-diversity region would be engaged by a few domains. The literature is replete with measures of diversity, particularly in biology and economics (e.g., Magurran, [155]). The Shannon diversity (or entropy), *H*, of a fingerprint was defined as [156]

$$H = -\sum_{i=1}^{S} p_i \ln p_i$$

where $S = 20$ was the number of task domains and $p_i$ corresponded to the $i^{th}$ domain proportion (for an improved Shannon index, see [157]). Diversity varied considerably across cortex (Figure 19), with "hot spots" apparent in dorsal-medial PFC, dorsal-lateral PFC, and anterior insula, among others; "cool spots" were observed in lateral temporal cortex, parts of posterior medial frontal/parietal cortex, and ventral-medial PFC/orbitofrontal cortex, among others.

--- Figure 18 ---

## 12. Comparing brain networks

The concept of functional fingerprint can be extended to networks. In this case, fingerprints can be computed by simultaneously considering all nodes within a network. Here, I illustrate the notion by using a version of network fingerprint that considers activations of all of the constituent regions of the network, a type of "union" operation (but note that several definition of network fingerprint are possible). Figure 20 shows the functional fingerprints of some of the networks discussed below. It is noteworthy, for instance, that the task-positive network and the task-negative network display fairly complementary profiles, matching the intuition that these networks are at times "anti-correlated"[158, 159].

--- Figure 20 ---

Multiple large-scale brain networks have been described, including general task-positive and task-negative networks, as well as more specific ones, such as "dorsal attention", "ventral attention", and "executive control" networks. Indeed, in recent years, interest in network science has led to the rapid growth of "new" networks. In several instances, investigators have proposed closely related networks (for instance, "dorsal attention" and "executive control"), prompting the possibility that they could be closely related, or possibly the same except for a chance in label. Thus, developing tools that help characterize and understand brain networks is of great relevance and could reveal principles of organization.

With this in mind, we asked the following question [153]: What is the relationship of the functions of regions belonging to a given network? One approach to address this question is to evaluate how homogeneous fingerprints are in a network. In other words, are fingerprints from the regions of network *X* more similar to each other than to those of regions from network *Y*? Our goal was not to investigate a unique set of networks, but instead consider possibly related (or even closely related) networks defined by different research groups and approaches, including meta-analysis, resting-state, and task-based approaches (see Table 1). To contrast brain networks to each other in terms of the functional fingerprints of the component regions, we employed a multivariate test based on "statistical energy" [160]. Briefly, statistical energy was used to evaluate whether two sets (that is, networks) of fingerprints were drawn from the same parent distribution. As described, the functional fingerprint for each constituent region within a network was represented as a 20-dimensional vector. The statistical energy metric compared distances of fingerprints within the same network relative to distances of fingerprints between different networks.

The statistical energy provided an estimate of the distance between two networks (based on the specific assignment of regions to networks). To characterize the magnitude of the distance, we employed a "percentile bootstrap method". Specifically, we determined the *null* ("chance") distribution of distances by randomizing the labels of the regions. That is, the assignment of a given region to a specific network $X$ or $Y$ was randomized. By repeating this process 10,000 times, the distributions shown in Figure 21 were generated (which corresponded to the set of potential differences between networks that one would obtain "by chance", namely without knowledge of the correct network affiliation of a given region). The final index of distance corresponded to the percentile of the actual observed distance within the bootstrapped distribution.

--- Figure 21 ---

Several network pairs were found to be fairly distinct (shown in red) or modestly distinct (shown in magenta). In Figure 21, each box shows the estimated "chance" distribution and the observed difference between two networks (blue vertical line): the more extreme the observed difference relative to the null distribution, the greater the difference between two networks. To index the strength of this difference, the percentile of the observed difference was determined relative to the null distribution. For instance, a clear difference was detected between FrontoParietalN vs. CinguloParietalN (see figure caption for explanation), task-positive and task-negative networks, respectively, defined via task co-activation data. The results demonstrate that common networks employed in the literature, in many instances, are distinct from one another under the task domain structure investigated here. Conversely, based on the functional repertoire of their components' nodes, some of the networks that have been distinguished from one another in the past are *not* strongly distinct.

**12.1 Are brain networks assortative?**

Assortativity refers to the tendency of "like to connect with like" (e.g., Christakis and Fowler [161]). In network research, assortative networks are those in which highly connected nodes are themselves preferentially connected to high-degree nodes [162]. Unfortunately, this type of description completely neglects node *function*. Thus, we evaluated assortativity by utilizing information contained in functional fingerprints.

Networks defined via task-based co-activation history, by construction, would be expected to exhibit some assortativity — as co-activation means that the regions of the network were engaged by the same task. However, it is less clear that the same would be observed for networks defined via resting-state data. But, because resting-state connectivity may reflect, at least in part, the co-activation history of two regions, resting-state networks likely display assortativity, too.

How does a specific network compare to a set of other networks in terms of the functional fingerprints of the constituent regions? Specifically, a positively assortative network would be one in which the distances within the network would be smaller than distances between regions of that network and regions outside that network. In other words, functional fingerprints within an assortative network would be relatively similar to each other and relatively dissimilar to fingerprints from other networks. To address this question, we again considered a set of nine networks and compared each network to the set of all "other" networks (Figure 22). In this analysis, three of the networks exhibited robust positive assortativity (here taken as values above the 95th percentile): FrontoParietalN, DorsalAttentionC, and FrontoParietalD (see figure caption for explanation)). Interestingly, one version of the task-negative network, CinguloParietalN, was somewhat *dis*-assortative. That is, its regions tended to be more dissimilar to each other that to those of other networks. These results are consistent with the notion that task-negative networks are relatively *heterogeneous*. Along these lines, recent work has suggested that task-negative networks can be subdivided in a number of ways [163].

--- Figure 22 ---

Together, our findings suggest that brain regions are very diverse functionally, in line with the points raised by Poldrack [10, 11]. The results indicate, too, that the operations of a given brain region can be understood in terms of its *functional fingerprint*, namely the task domains that systematically engage the region. Beyond the descriptive aspects of the approach, it outlines a framework in which a region's function is viewed as inherently *multidimensional*: a vector defines the fingerprint of a region in the context of a specific domain structure. Although the domain that we explored used a task classification scheme from an existing database, it was not the only one possible. How should one define the domain structure? One hope is that cognitive ontologies can be defined that meaningfully carve the "mental" into stable categories [8, 164]. I contend, however, that *no single ontology will be sufficient*. Instead, it is better to conceive of several task domains that are useful and complementary in characterizing brain function and/or behavior. Thus, a region's functional fingerprint needs to be understood in terms of a *family of (possibly related) domains*. Finally, the framework can be extended to networks, provides a way to compare them, and to advance our understanding of the properties of constituent nodes.

**13. Understanding a region's function by exploring its co-activation partners**

The results shown in Figure 19 suggest that many brain regions are functionally heterogeneous. Here, we illustrate a strategy that we and others have developed to study a region's contributions to brain function and behavior. In a nutshell, in addition to studying what a brain region does, study what its *functional partners* do. To illustrate the idea, consider the insular cortex, a functionally diverse brain region that has been characterized by its involvement in somatic and visceral sensory processes, autonomic regulation, as well as motor processing [165]. Earlier views of the insula as primarily a low-level, emotion-related structure have given way to more complex and multifaceted views of insular function in recent years.

Again, we took advantage of the existence of large databases of functional neuroimaging studies to perform, in this case, *co-activation* meta-analysis (see [118]). Task-based co-activation analysis was conducted to capture the tendency of insular subdivisions to be active together with other brain areas during the same experimental task. Each of three insular subdivisions (on two hemispheres) served as a seed region, and co-activation was determined by moving a searchlight across the brain in a voxel-wise manner (to account for the influence of other insular subdivisions, partial correlations were computed). For each insular subregion, extensive co-activation was observed; notably, co-activation overlap was extensive (Figure 23). Next, we determined the functional fingerprint of the co-activating partners of each insular subdivision. For example, for the left dorsal anterior insula, we considered activations in all voxels shown in Figure 24A, as if they were a single "region." Figure 24B displays the "common" fingerprint for all the insular subdivisions. As can be seen, the common fingerprint was functionally diverse and encompassed all task domains probed. Furthermore, to understand what each insular subdivision expressed to a greater extent relative to their mean, we also determined profiles for each insular subregion by first subtracting a "mean fingerprint".

--- Figures 23 and 24 ---

The approach illustrated here complements the standard way that a brain region is studied. By investigating the functions of a region's functional partners the goal is to better understand the repertoire of functions that the region of interest participates in. This way, a broader characterization of the structure-function mapping is obtained. For related approaches, see [166-168].

## 14. Conclusions

In this paper, I discussed several aspects that inform the mapping between structure and function in the brain. As others in the past, I argue that a network perspective should supplant the common strategy of understanding the brain in terms of individual regions. Whereas this perspective is needed for a fuller characterization of the mind-brain, it should not be viewed as panacea. For one, the challenges posed by the many-to-many mapping between regions and functions is not dissolved by the network perspective. In a sense, the problem is ameliorated, but clearly one should *not* anticipate a *one*-to-*one* mapping when the network approach is adopted. Furthermore, decomposition of the brain network in terms of meaningful clusters of regions, such as the ones generated by community-finding algorithms, does not by itself reveal "true" subnetworks. Given the hierarchical and multi-relational relationship between regions, multiple decompositions will offer different "slices" of a broader landscape of networks within the brain. Finally, I described how the function of brain regions can be characterized in a multidimensional manner via the idea of diversity profiles. The concept can also be used to describe the way different brain regions participate in networks.

**Table 1**
Network definitions.

| Network | Abbrevi... |
|---|---|
| Fronto-parietal, seeded from the left intraparietal sulcus as in (Toro et al., 2008) | FrontPar |
| Cingulo-parietal, seeded from the anterior cingulate cortex as in (Toro et al., 2008) | CingPar |
| Dorsal Attention (Yeo et al., 2011) | DorsAtt |
| Ventral Attention (Yeo et al., 2011) | VentAtt |
| Control (Yeo et al., 2011) | Control |
| Default (Yeo et al., 2011) | Default |
| Fronto-parietal (Dosenbach, et al., 2007) | FrontPar |
| Cingulo-opercular (Dosenbach et al., 2007) | CingOpe |

**Figure Captions**

Figure 1. Structure-function mapping in the brain. A central argument of the paper is that because the mapping from structure to function is *many*-to-*many*, understanding the instantiation of functions by the brain necessitates sophisticated frameworks whose basic elements are networks, not regions. Abbreviations: A1, …, A4: areas 1 to 4; amyg: amygdala; F1, …, F4: functions 1 to 4; Reproduced with permission [2].

Figure 2. Whole-brain network connectivity structure. (A) The analysis considered existing anatomical connectivity data of an extensive set of cortical and subcortical areas spanning most major brain sectors. The basal ganglia refer to nuclei at the base of the forebrain, including the amygdala. (B) Innermost "core" circuit. Notably, several amygdala nuclei were included in the inner core. Reproduced with permission [34].

Figure 3. Structure-function mapping and networks. (A) The "landscape of behavior" displays a caricature of the multidimensional space of behaviors. Abbreviations: $A_1$, $A_2$, $A_N$, $B_1$, and $B_N$: brain regions; $N_1$ and $N_2$: networks; $P_i$ and $P_j$: processes. (B) Intersecting networks. The networks $C_K$ and $C_L$ (and the additional ones) intersect at node $A_n$. (C) Dynamic aspects. Region $A_N$ will have network affiliations that vary as a function of time. Therefore, the processes carried out by the emerging networks will evolve across time and lead to dynamic "landscapes of behavior". The four time points represented are such $t_1 \approx t_2$ but far from $t_3 \approx t_4$. (D) Structure-function mappings in the case of networks. Two networks may instantiate similar processes, a case of many-to-one mapping. The reverse relationship is also suggested to apply to networks, namely, one-to-many mappings (see text). Reproduced with permission [2].

Figure 4. Network interactions. The same regions $R_i$ may comprise distinct networks depending on how the regions interact both in terms of strength-across-time (top row) and time (bottom row). For instance, when the $R_1 \rightarrow R_2$ link is strong, network $N_1$ will behave differently from when the $R_2 \rightarrow R_1$ link is strong (in that case, the network is being labeled $N_2$). This could occur, for instance, due to plasticity. The bottom row illustrates that inter-region interactions may follow a different temporal order, thereby leading to different function (in this case, the networks were labeled $N_1$ and $N_2$). In both cases, the same network label $N_1$ could have been used for the two scenarios, with the understanding that $N_1$ varies as a function of time. Reproduced with permission [2].

Figure 5. Network structure and *hub* nodes. Nodes with unusually high connectivity may be considered "hubs", and are likely to have important roles in determining information flow. Reproduced with permission [2].

Figure 6. Network associated with Brodmann's area 46 in dorsal-lateral PFC. Given its high connectivity, the area can be considered a hub. Reproduced with permission [65].

Figure 7. Conventional and alternative views of thalamo-cortical circuits. In the conventional view, cortical communication is accomplished via pathways between cortical sites. In the alternative view, as proposed by Sherman and colleagues, higher-order thalamic nuclei play a prominent role in this communication, and direct cortico-cortical pathways may be less important. FO, first order; HO, higher order. Reproduced with permission [74].

Figure 8. Prefrontal cortex connectivity. Fraction of frontal areas that receive signals from each modality as a function of the number of connectivity "steps" within frontal cortex. Zero indicates the areas that receive a direct projection from the indicated sensory or motor modality, and one indicates the fraction of areas that would receive the signal after a single step within frontal cortex. Amy, amygdala; Aud, auditory; G/O, gustatory/olfactory; Hip, hippocampus; Mot, motor; MS, multisensory; SS, somatosensory; Vis, visual. Reproduced with permission [38].

Figure 9. Hypothalamic ascending connectivity. Summary of the four major pathways from the hypothalamus to cerebral cortex schematized on a flattened representation of the rat brain. The basal ganglia here refer to the basal forebrain and the amygdala complex. Note that one of the indirect connections first descends to the brainstem. BG, basal ganglia; BS, brainstem; CTX, cortex; HY, hypothalamus; TH, thalamus. Reproduced with permission [29].

Figure 10. Basic plan of the vertebrate brain. From Striedter (2005) with permission.

Figure 11. Basic plan of the vertebrate brain represented as a "flat" map. Reproduced from Swanson (2007) with permission.

Figure 12. Summary representation of the amygdala in rats (A), monkeys (B), and humans (C). Glial cells, circles; neurons, triangles: rats < monkeys < humans. Neuron density: rats > monkeys > humans. Glia density: rats ¼ monkeys ¼ humans. Glia/neuron ratio: rats < monkeys < humans. Connectivity with visceral and autonomic systems (mainly via the central nucleus): rats ¼ monkeys ¼ humans. Connectivity with cortical systems (between the neocortex and the lateral, basal, and accessory basal nuclei): rats < monkeys < humans. The proportions of the different parameters are not precisely scaled. Reproduced with permission [88].

Figure 13. Function and structure. The relationship between structure and function can be nuanced and complex. While panel (A) describes the "default" case of a structural connection leading to a functional relationship, the other panels describe how the link can be far from straightforward. See text for further discussion. Abbreviations: $R_i$: regions; $C_i$: contexts. Reproduced with permission [2].

Figure 14. Emotion alters the response pattern across early visual cortex. (A) Both fearful and neutral (not shown) faces were presented with a ring that indicated whether it was a safe or threat (possible mild shock) condition. (B) Functional connectivity between visual regions V1, V2, V3, and V4 during affective (left) and neutral (right) context as indexed via pairwise correlations (average correlation shown). (C) Correlation matrix of differential responses for fearful vs. neutral faces in both affective and neutral contexts (ring colors). The upper-left part is the same as in panel B. Of the entire matrix, that was the only part that was robustly altered as a function of context. Abbreviations: Varea: correlations within early visual areas; Oarea: correlations within "other areas"; VOarea: correlations between early visual and "other areas". L: left; R: right; PrCu: precuneus; STS: superior temporal sulcus; IPS: intraparietal sulcus; pSMA: pre-supplementarymotor area; PrCs: precentral sulcus; FEF: frontal eye field; INS: insula. Colors code for correlation values as indicated by the color scale. Adapted with permission [105].

Figure 15. The effects of common efferents, two-step relay, and common afferents on functional connectivity. *$P< 0.05$, Tukey test. Reproduced with permission [114].

Figure 16. Network structure and reward. (A) Community detection was applied to the set of brain regions that responded more strongly to reward vs. no-reward at the cue phase (see Figure 6.4). Two communities were detected; please see [123] for region abbreviations. (B) Comparison of the pattern of connectivity between reward and no-reward contexts revealed increases during the former. The increases were observed mostly *between* the two communities, reflecting increased integration with reward. The polar plot shows increases in functional connectivity of the right caudate with nearly all regions belonging to the "other" community. Line width represents the relative strength of the functional connectivity between regions. Adapted with permission [123].

Figure 17. Network structure and threat. (A) Community detection was applied to the set of brain regions that responded more strongly to threat vs. safe at the cue phase (see Figure 5.7). Two communities were detected; please see [123] for region abbreviations. (B) Changes in threat versus safe connectivity for all pairs of regions within the community on the left in panel A. Dark colors indicate no change; warm colors indicate threat greater than safe; cool colors indicate threat smaller than safe. One of the effects of threat may have been to "disconnect" cortical regions from each other, possibly leading to performance impairments. Adapted with permission [123].

Figure 18. Functional fingerprints of regions and networks. (A) The polar plots illustrate the fingerprints of three brain regions. Each vertex corresponds to one of the domains investigated. Both the left anterior insula and the left intraparietal sulcus exhibited diverse functional profiles. The superior temporal gyrus in the vicinity of auditory cortex was less diverse, though the fingerprint revealed its involvement in emotional processing, in addition to audition. (B) The polar plots illustrate the fingerprints of two brain networks, which were defined by Toro et al. [118] based on a meta-analysis of task activation data. The frontal-parietal "attention" network was a task-*positive* network generated by "seeding" the left intraparietal sulcus. The cingulate-parietal "resting-state" network was a task-negative network generated by "seeding" the ventral-anterior medial PFC. Although both networks are quite diverse, the analysis revealed that they are fairly complementary to one another. Reproduced with permission [153].

Figure 19. Diversity map. (A) Areas of higher diversity are shown in warm colors and areas of lower diversity are shown in cool colors (color bar represents $H$ values). Locations without colors did not have enough studies for the estimation of diversity. Reproduced with permission [153].

Figure 20. Network functional fingerprints. Reproduced with permission [153].

Figure 21. Network comparison. A multivariate comparison with permutation testing was used to compare pairs of networks. The distributions portray the null distribution of possible differences between each pair. The blue vertical bars indicate the observed difference, which is shown on top of each box in terms of its percentile relative to the null distribution (when not shown, the bar was located to the right of the displayed area). For illustration, comparisons with percentiles>95% are shown in red and comparisons with percentiles> 90% are shown in magenta. For example: FrontoParietalN and CinguloParietalN were very different, DorsalAttentionC and VentralAttentionC were distinct but to a lesser extent, and CinguloParietalN and DefaultC were similar. See Table 1 for network explanations. Reproduced with permission [153].

Figure 22. Network assortativity. Assortativity measures the extent to which functional fingerprints from regions of the same network are more similar to each other than to fingerprints from other networks. The percentile scores provide an indication of the degree of assortativity (or dis-assortativity in the case of CinguloParietalN). See Table 1 for network explanations. Reproduced with permission [153].

Figure 23. Overlap between connection partners of each insular subdivision. To facilitate displaying overlap, the corresponding right and left insular subregions were pooled together resulting in three

insular subregions (dorsal anterior, ventral anterior, posterior insula). Voxels shown in green-to-red colors were coactive with two of the three subregions (the color bar indicates the strength of overlap, specifically, the smallest value of the two strongest partial correlations). Voxels in blue were coactive across all three subdivisions. Adapted with permission [169].

Figure 24. (A) Coactivation of insula subdivisions. Using data from the Neurosynth database, task-based coactivation profiles were determined for each insular subdivision by moving a searchlight in a voxel-wise manner. The color bar indicates the partial correlation value with the specific insular subregion "seed" when all other subdivisions were also considered. (B) "Common" functional fingerprint of insular subdivisions. The common fingerprint was determined by combining all six insular subregion (see text). All task domains were engaged by each subregion at least some of the time. TOM = theory of mind; MemWork = working memory; MemOther = long-term memory. Adapted with permission [169].

Brain networks[57] Buckner RL, Sepulcre J, Talukdar T, Krienen FM, Liu H, Hedden T, et al. Cortical hubs revealed by intrinsic functional connectivity: mapping, assessment of stability, and relation to Alzheimer's disease. Journal of Neuroscience. 2009;29:1860-73.

[58] Cacioppo JT, Tassinary LG. Inferring psychological significance from physiological signals. American Psychologist. 1990;45:16-28.

[59] Marder E, Goaillard JM. Variability, compensation and homeostasis in neuron and network function. Nat Rev Neurosci. 2006;7:563-74.

[60] Lashley KS. Basic neural mechanisms in behavior. Psychological review. 1930;37:1-24.

[61] Passingham RE, Stephan KE, Kotter R. The anatomical basis of functional localization in the cortex. Nature Reviews Neuroscience. 2002;3:606-16.

[62] Newman M. Networks: An Introduction. New York City: Oxford University Press; 2010.

[63] Guimera R, Nunes Amaral LA. Functional cartography of complex metabolic networks. Nature. 2005;433:895-900.

[64] Guimera R, Sales-Pardo M, Amaral LAN. Classes of complex networks defined by role-to-role connectivity profiles. Nature Physics. 2007;3:63-9.

[65] Sporns O, Honey CJ, Kotter R. Identification and classification of hubs in brain networks. PLoS ONE. 2007;2:e1049. doi:10.371/journal.pone.00001049.

[66] Rubinov M, Sporns O. Complex network measures of brain connectivity: uses and interpretations. NeuroImage. 2010;52:1059-69.

[67] Alexander GE, DeLong MR, Strick PL. Parallel organization of functionally segregated circuits linking basal ganglia and cortex. Annual review of neuroscience. 1986;9:357-81.

[68] Chalfin BP, Cheung DT, Muniz JA, de Lima Silveira LC, Finlay BL. Scaling of neuron number and volume of the pulvinar complex in New World primates: comparisons with humans, other primates, and mammals. Journal of Comparative Neurology. 2007;504:265-74.

[69] Grieve KL, Acuna C, Cudeiro J. The primate pulvinar nuclei: vision and action. Trends in neurosciences. 2000;23:35-9.

[70] Shipp S. The functional logic of cortico-pulvinar connections. Philosophical Transactions of the Royal Society of London B: Biological Sciences. 2003;358:1605-24.

[71] Stepniewska I. The pulvinar complex. In: Kaas J, Collins CE, editors. The primate visual system. Boca Raton, Florida: CRC Press; 2004. p. pp 53-80.

[72] Romanski LM, Giguere M, Bates JF, Goldman-Rakic PS. Topographic organization of medial pulvinar connections with the prefrontal cortex in the rhesus monkey. Journal of Comparative Neurology. 1997;379:313-32.

[73] Barbas H, Henion TH, Dermon CR. Diverse thalamic projections to the prefrontal cortex in the rhesus monkey. Journal of Comparative Neurology. 1991;313:65-94.

[74] Sherman SM. The thalamus is more than just a relay. Current Opinions in Neurobiology. 2007;17:417-22.

[75] Theyel BB, Llano DA, Sherman SM. The corticothalamocortical circuit drives higher-order cortex in the mouse. Nature Neuroscience. 2010;13:84-8.

[76] Saalmann YB, Pinsk MA, Wang L, Li X, Kastner S. The pulvinar regulates information transmission between cortical areas based on attention demands. Science (New York, NY. 2012;337:753-6.

[77] Swanson LW. The amygdala and its place in the cerebral hemisphere. Annals of the New York Academy of Sciences. 2003;985:174-84.

[78] Amaral DG, Price JL, Pitkanen A, Carmichael ST. Anatomical organization of the primate amygdaloid complex. In: Aggleton J, editor. The Amygdala: neurobiological aspects of emotion, memory, and mental dysfunction. New York: Wiley-Liss; 1992. p. 1-66.

[79] Sah P, Faber E, De Armentia ML, Power J. The amygdaloid complex: anatomy and physiology. Physiological reviews. 2003;83:803-34.

[80] Amaral DG, Price JL. Amygdalo-cortical projections in the monkey (Macaca fascicularis). Journal of Comparative Neurology. 1984;230:465-96.

[81] Ghashghaei HT, Hilgetag CC, Barbas H. Sequence of information processing for emotions based on the anatomic dialogue between prefrontal cortex and amygdala. NeuroImage. 2007;34:905-23.

[82] Messinger A, Winkle CC, Tootell RB, Ungerleider LG. Using electrical stimulation and fMRI to map connections of the amygdala in monkey. 2011 Society for Neuroscience. Washington, D.C.2011.

[83] Swanson LW. Cerebral hemisphere regulation of motivated behavior. Brain research. 2000;886:113-64.
33

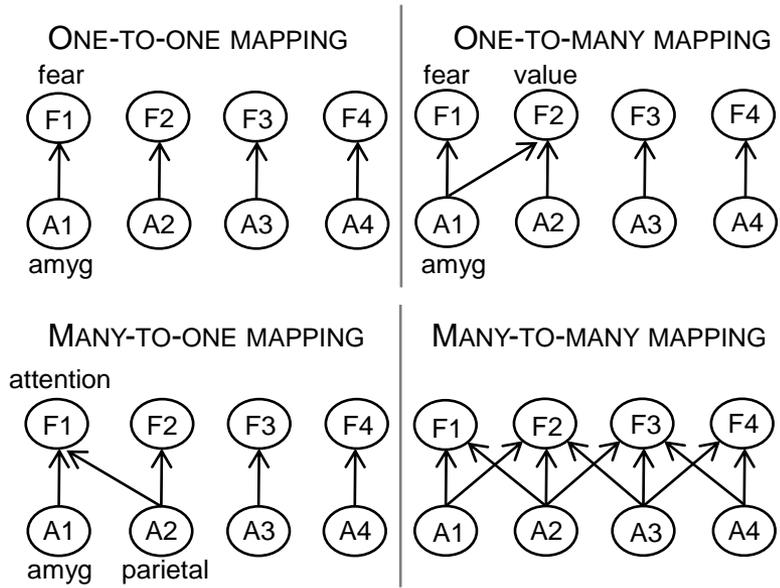

Figure 1

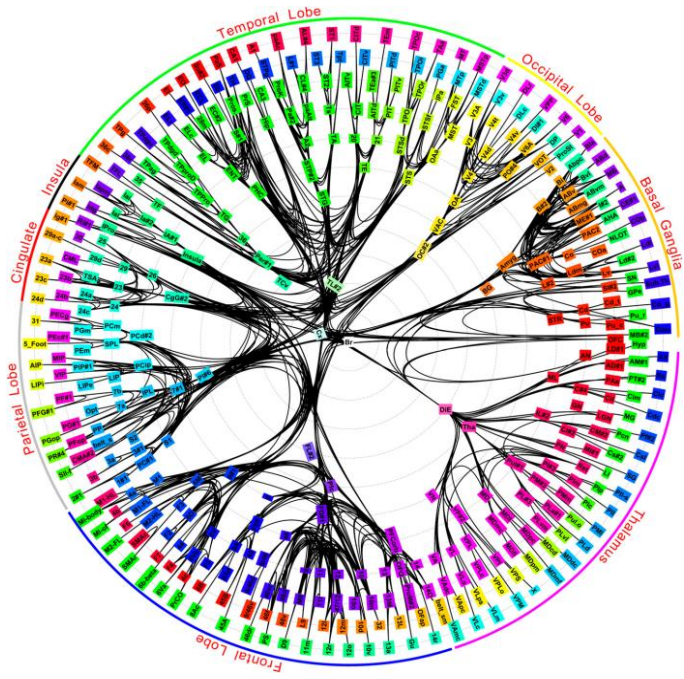

Figure 2

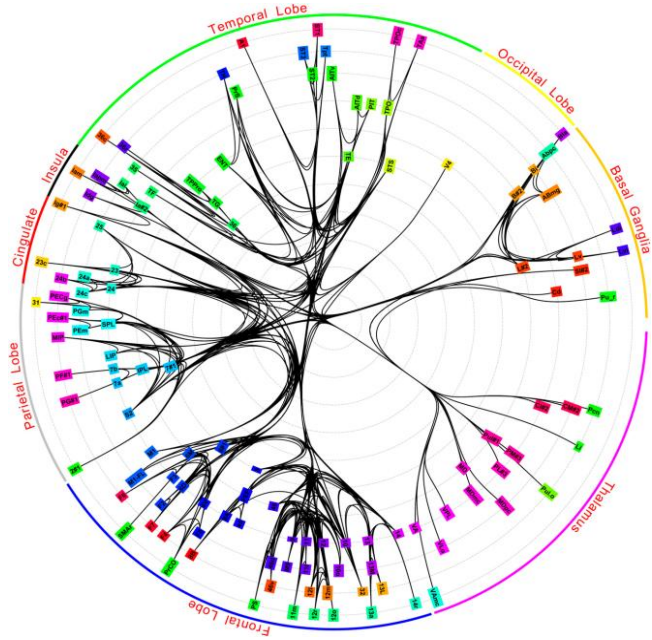

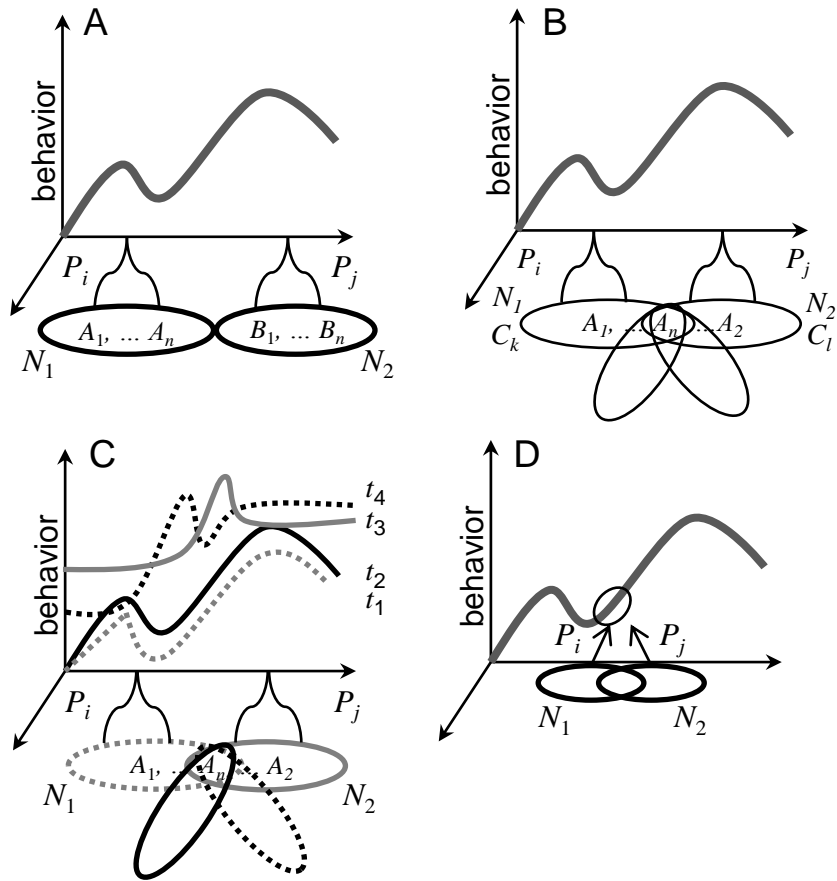

Figure 3

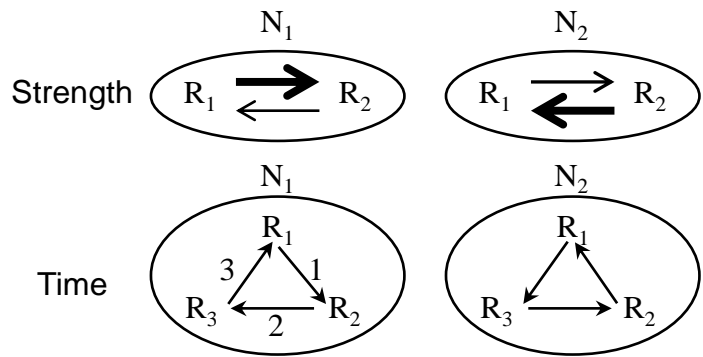

Figure 4

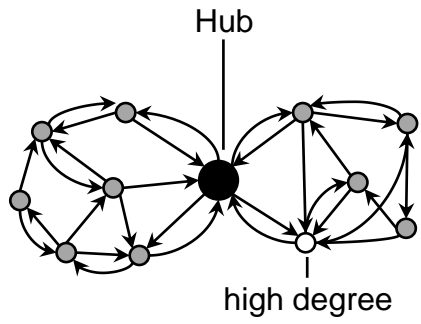

Figure 5

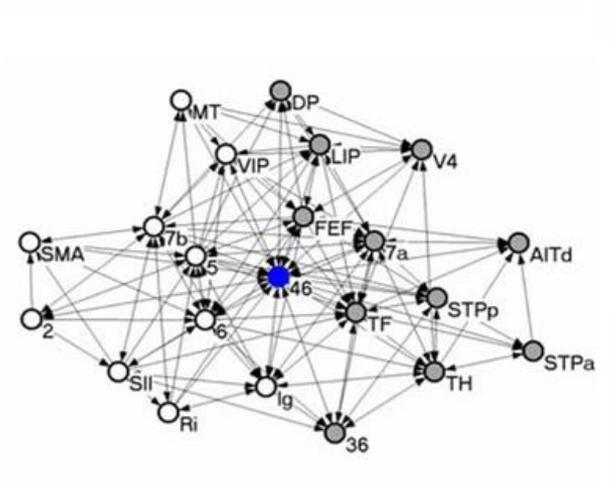

Figure 6

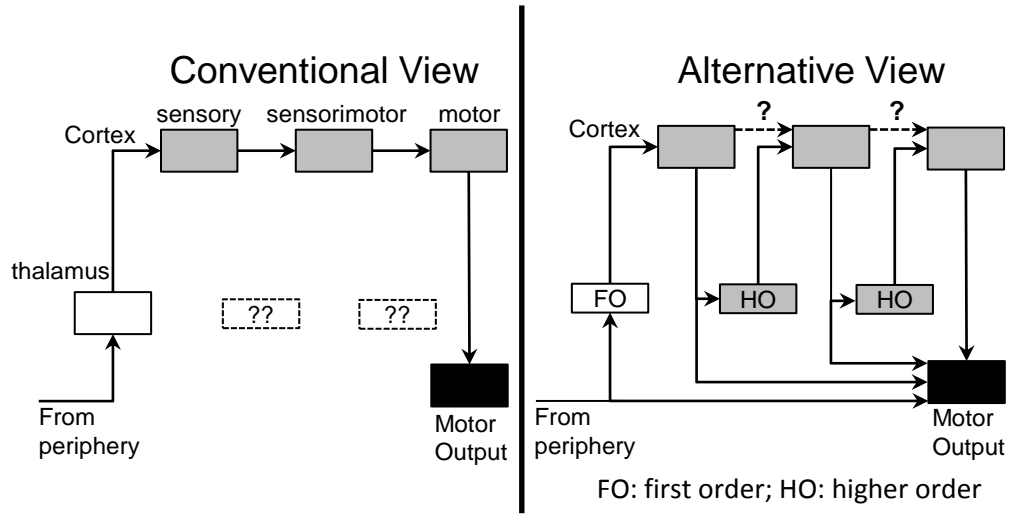

Figure 7

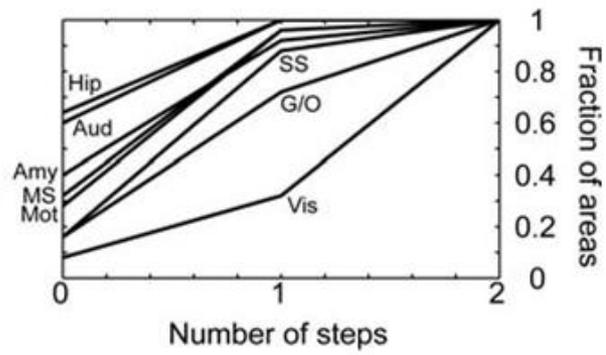

Figure 8

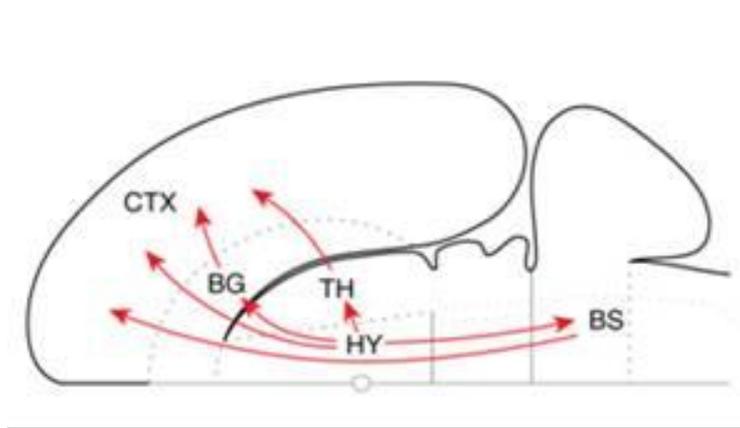

Figure 9

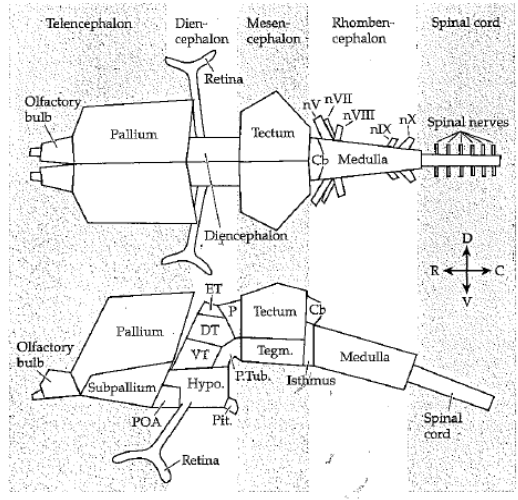

Figure 10

**Figure 3.8 The Vertebrate Brain Archetype** Schematic drawings of the vertebrate brain region archetype from dorsal (top) and lateral (bottom) perspectives. Some elements have been omitted for the sake of clarity. Abbreviations: C = caudal; Cb = cerebellum; D = dorsal; Dienc. = diencephalon; DT = dorsal thalamus; ET = epithalamus; Hypo. = hypothalamus; nV = trigeminal nerve; nVII = facial nerve; nVIII = octaval nerve; nIX = glossopharyngeal nerve; nX = vagal nerve; P = pretectum; Pit. = posterior pituitary; POA = preoptic area; P. Tub. = posterior tuberculum; R = rostral; Tegm. = tegmentum; V = ventral; VT = ventral thalamus.

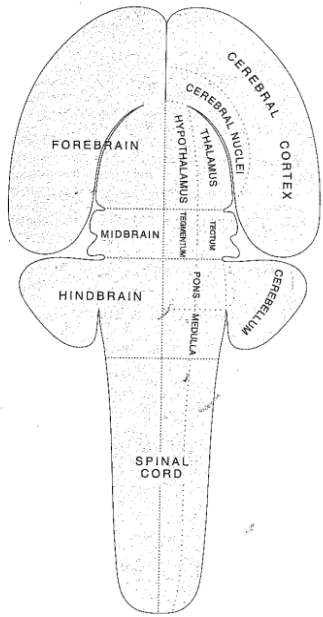

Figure 11

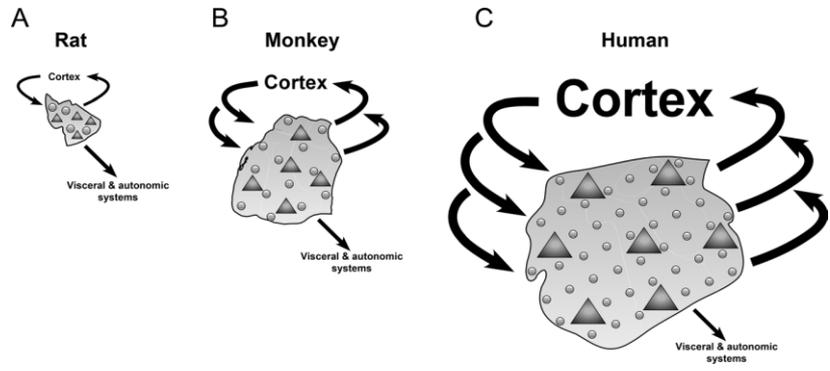

Figure 12

(A) 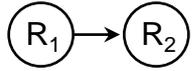

(B) 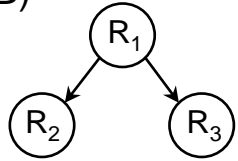

(C) 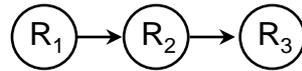

(D) 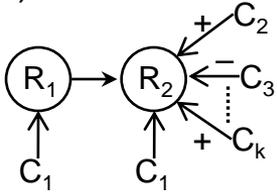

(E) 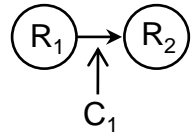

Figure 13

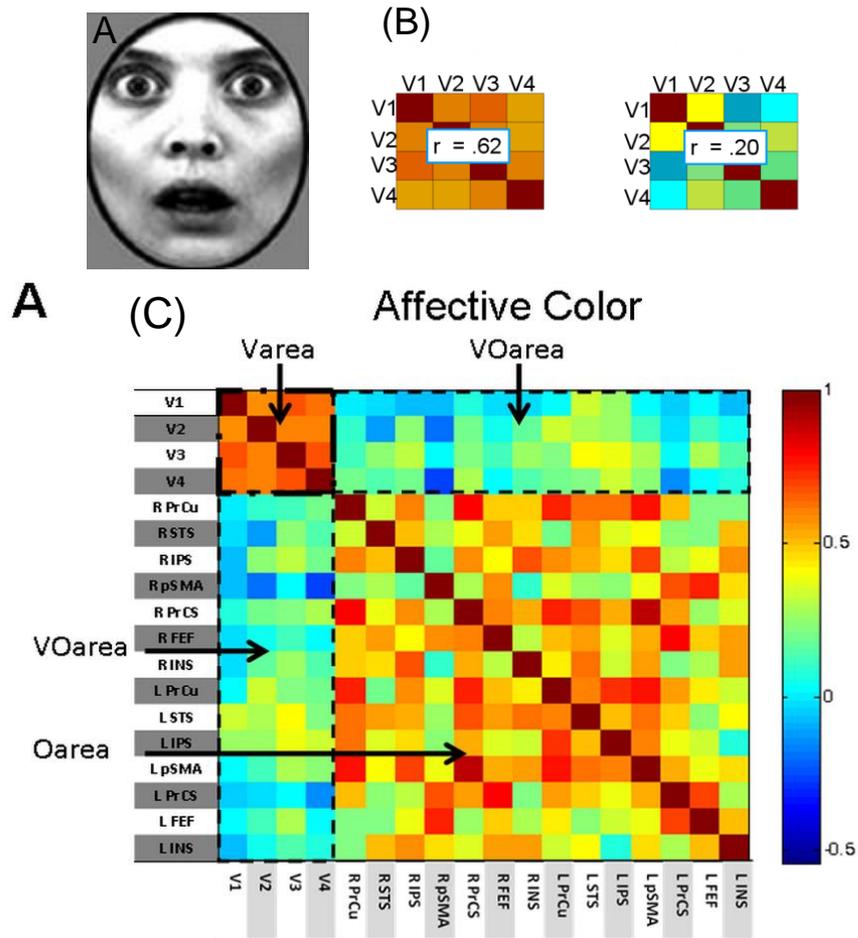

Figure 14

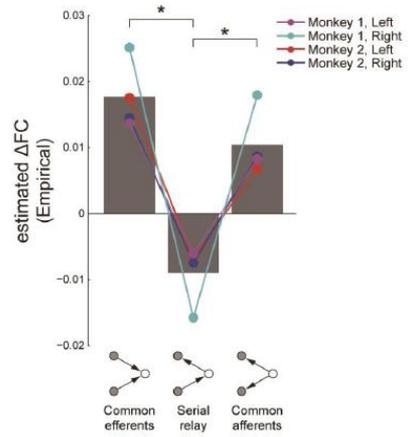

Figure 15

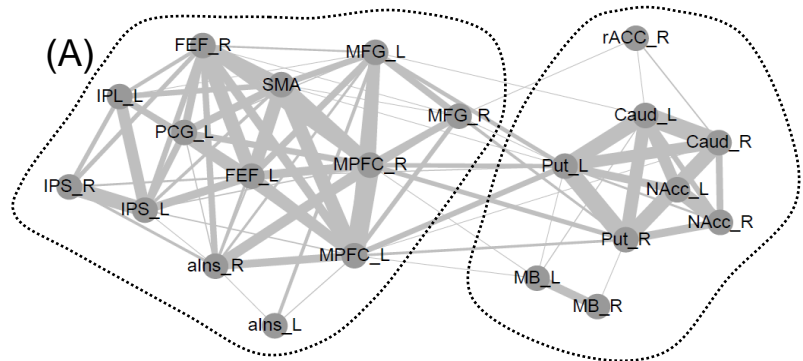
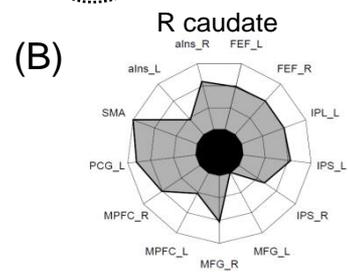

Figure 16

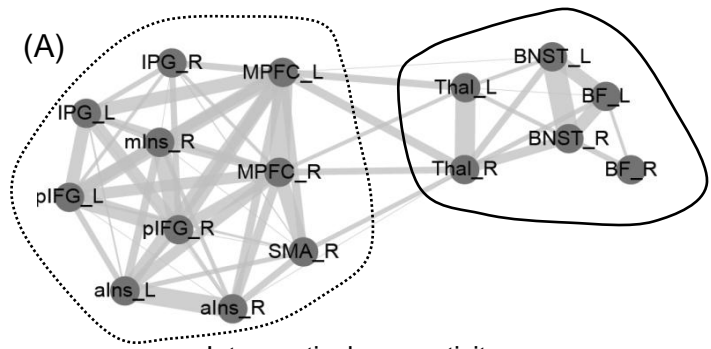

Intra-cortical connectivity

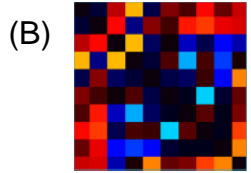

Figure 17

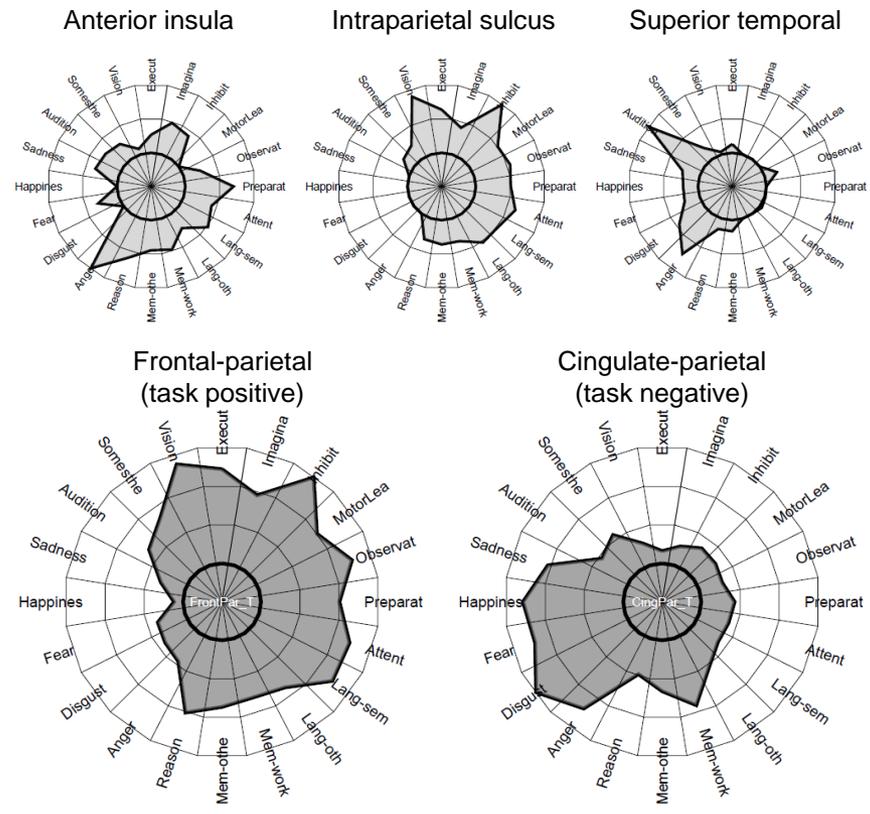

Figure 18

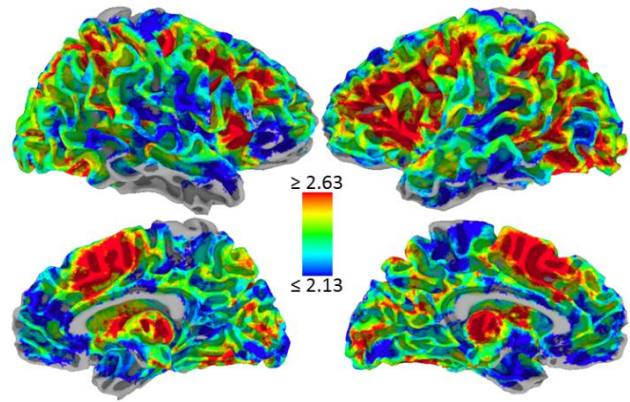

Figure 19

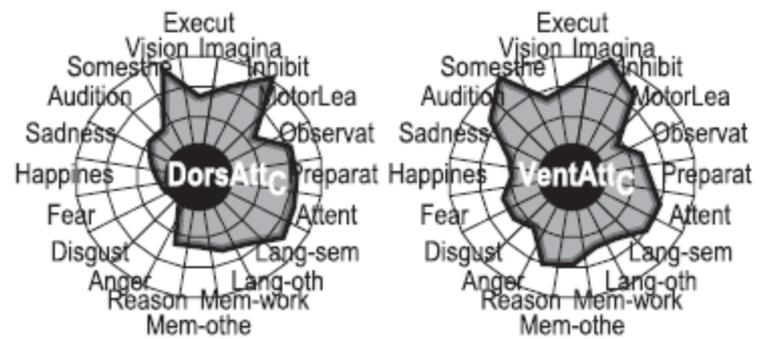

Figure 20

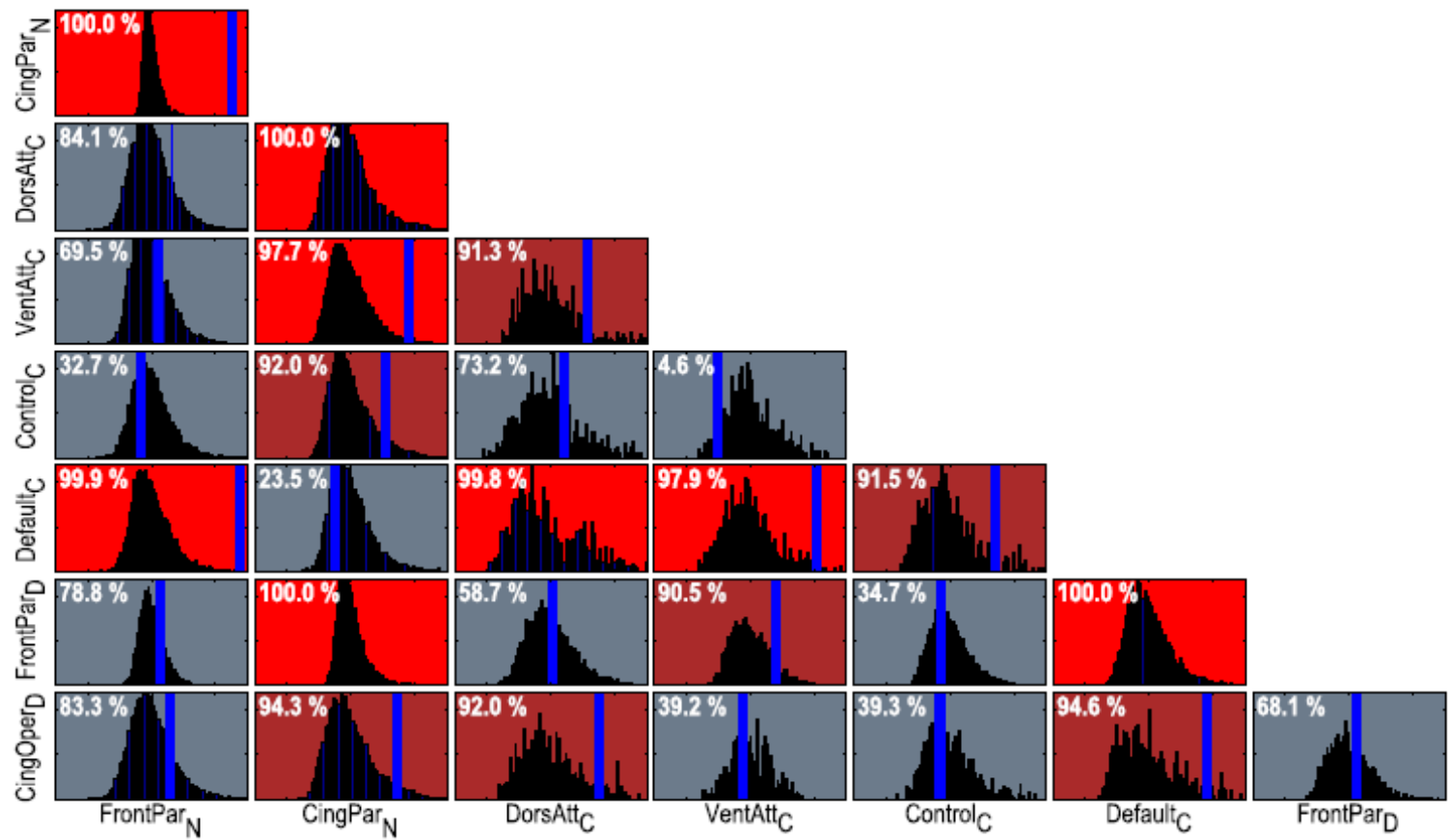

Figure 21

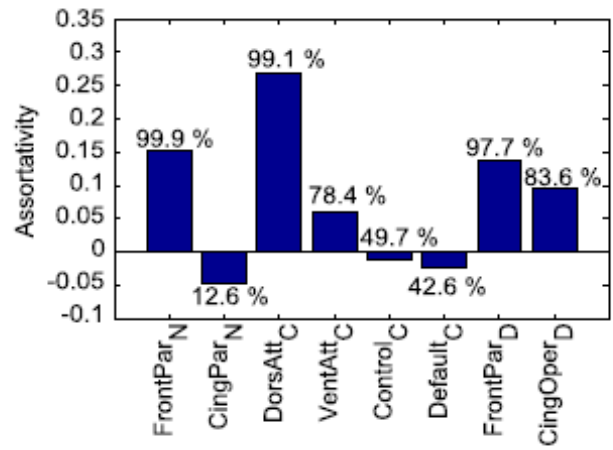

Figure 22

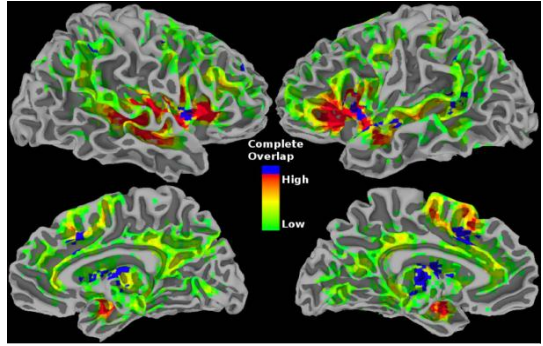

Figure 23

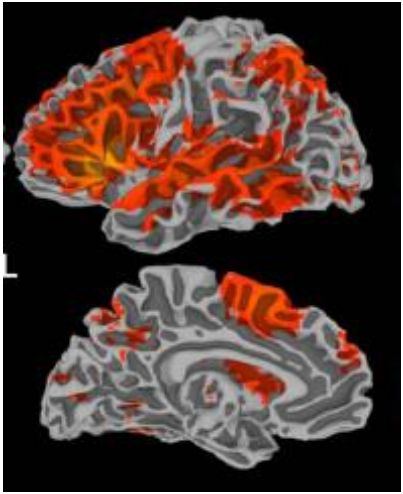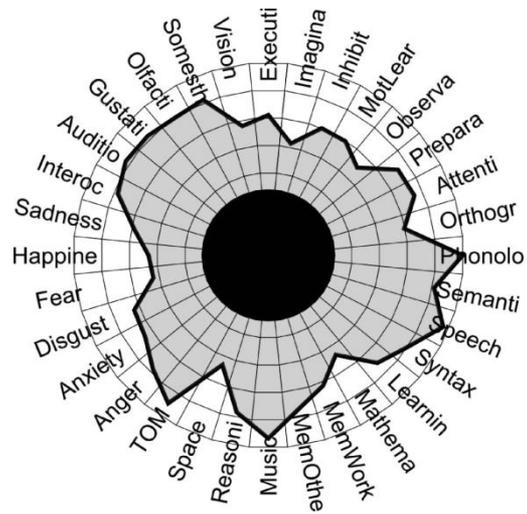

Figure 24